\documentclass[a4paper,11pt]{article}

\usepackage{jheppub}
\usepackage[utf8]{inputenc}
\usepackage{graphicx,cancel,float}
\usepackage{amssymb}
\usepackage{textcomp}
\usepackage{amsmath}
\usepackage{tabularx}
\usepackage{bm}
\usepackage{times}
\usepackage{color}
\usepackage{slashed}
\usepackage{multirow}
\usepackage{verbatim}
\usepackage{epsfig,epstopdf}
\usepackage{subfigure}
\allowdisplaybreaks[4]

\def\lsim{\mathrel{\raise.3ex\hbox{$<$\kern-.75em\lower1ex\hbox{$\sim$}}}}
\def\gsim{\mathrel{\raise.3ex\hbox{$>$\kern-.75em\lower1ex\hbox{$\sim$}}}}

\definecolor{orange}{rgb}{1,0.5,0}

\newcommand{\bew}{\begin{widetext}}
\newcommand{\enw}{\end{widetext}}
\newcommand{\bee}{\begin{equation}}
\newcommand{\ene}{\end{equation}}
\newcommand{\bea}{\begin{eqnarray}}
\newcommand{\ena}{\end{eqnarray}}
\newcommand{\bes}{\begin{subequations}}
\newcommand{\ens}{\end{subequations}}

\def\cald{\mathcal{D}}

\def\calj{\mathcal{J}}

\def\calm{\mathcal{M}}

\def\calo{\mathcal{O}}
\def\calp{\mathcal{P}}

\subheader{\it \today}

\title{\Large{\bf Laser induced Compton Scattering to Dark Matter in Effective Field Theory}}

\author[a]{Kai Ma}
\emailAdd{kai@xauat.edu.cn}
\author[b]{Tong Li}
\emailAdd{litong@nankai.edu.cn}

\affiliation[a]{Faculty of Science, Xi'an University of Architecture and Technology, Xi'an, 710055, China}
\affiliation[b]{School of Physics, Nankai University, 94 Weijin Road, Tianjin 300071, China}

\abstract{
The detection of light dark matter (DM) is a longstanding challenge in terrestrial experiments. High-intensity facilities with intense electromagnetic field may provide a plausible strategy to study strong-field particle physics and search for light DM.
In this work, we propose to search for light DM particles through nonlinear Compton scattering in the presence of a high-intensity laser field. An ultra-relativistic electron beam collides with an intense laser pulse containing a number of optical photons and then decays to a pair of DM particles. We take into account the Dirac-type fermionic DM in a leptophilic scenario and the DM-electron interactions in the framework of effective field theory. The decay rates of an electron to a DM pair are calculated for effective DM operators of different bilinear products. We show the sensitivities of laser induced Compton scattering to the effective cutoff scale for DM lighter than 1 MeV and compare them with direct detection experiments.
}

\begin{document}

\maketitle
\setcounter{page}{2}

\newpage

\section{Introduction}
\label{sec:Intro}

A number of cosmological and astrophysical observations clearly support the existence of dark matter (DM) in the universe~\cite{Bertone:2004pz,Young:2016ala,Arbey:2021gdg}.
While it is motivated by its gravitational effects,
the DM can also be manifested by its weak but non-vanishing couplings to the Standard Model (SM) particles~\cite{Leane:2020liq,Slatyer:2021qgc}.
Because of the sizable relic abundance of DM,
its annihilation in high-density regions of the universe can indirectly induce
observational anomalous signals of leptons or photons~\cite{Bouquet:1989sr,Baltz:2002we,John:2021ugy,Bi:2009uj,Ibarra:2009bm}.
Various detection strategies have also been proposed to search for DM in direct detection~\cite{Goodman:1984dc} and collider experiments~\cite{Bai:2010hh,Goodman:2010yf,Goodman:2010ku}. However, the microscopic properties of DM still remain unknown, and the DM particle has not been observed in the terrestrial facilities. The direct detection experiments are normally not sensitive to DM lighter than 1 MeV due to the limitation of recoil energy threshold.
The high-intensity facility with an intense electromagnetic field may compensate for the shortcomings of other experiments for the detection of new physics beyond the SM~\cite{Wistisen:2020czu,Ouhammou:2022jys,Tiedau:2024obk,Dillon:2018ypt,King:2018qbq,Dillon:2018ouq,Bai:2021gbm,King:2019cpj,Beyer:2021mzq,Ma:2024ywm}.

The laser with an intense electromagnetic field strength provides an essential tool for exploring strong-field particle physics in the high-intensity frontier.
In 1951, J. Schwinger showed that at a field strength of $E=m_e^2/e\sim 1.32\times 10^{18}~{\rm V/m}$, the vacuum of quantum electrodynamics (QED) becomes unstable and can decay into a pair of electron and positron~\cite{Schwinger:1951nm}. This so-called Schwinger field is of particular interest because it exhibits non-perturbative QED. The vacuum decay probability yields zero to all orders of perturbative theory. In 1990s, the SLAC experiments observed two strong-field processes through the interaction of
an ultra-relativistic electron beam with a terawatt laser pulse~\cite{Burke:1997ew,Bamber:1999zt}. Under the illumination of a high-intensity laser field, an electron can absorb the energy of multiple laser photons and then decay into a high-energy photon. This high-energy photon can further collide with the laser photons and decay into an electron-positron pair. These processes are known as nonlinear Compton scattering
\begin{eqnarray}
e^- + {\rm laser} \to e^- + \gamma\;,
\end{eqnarray}
and nonlinear Breit-Wheeler pair production
\begin{eqnarray}
\gamma+ {\rm laser} \to e^+ + e^-\;.
\end{eqnarray}
The precision measurement of these two processes has catalyzed the studies of strong-field QED and nonlinear QED (see Greiner et al.'s textbook~\cite{Greiner:1992bv} and recent reviews in Refs.~\cite{Hartin:2018egj,Fedotov:2022ely}).
When the strength of the laser's electromagnetic plane wave is sufficiently intense,
the nonlinear contributions from the higher-order terms in the scattering process become important. The benefit of absorbing more than one photon is that the resultant scattering cross section is significantly larger (in unit of barn) than those of conventional QED processes in beam collisions.

In this work, we propose to search for DM using the laser-assisted Compton scattering process. A high-energy electron beam collides with an intense laser pulse containing a number of optical photons and then decays into a pair of DM particles
\begin{eqnarray}
e^- + {\rm laser} \to e^- + \chi + \overline{\chi}\;,
\end{eqnarray}
where we assume Dirac-type fermionic DM denoted by $\chi$ and its anti-particle $\overline{\chi}$.
Unlike in the QED Compton scattering, the final states in our case are composed of one single visible electron and the missing transverse momentum carried away by the DM pair. It is analogous to the characteristic mono-electron signature at colliders~\cite{Bai:2012xg,Ma:2024aoc,ATLAS:2014wra,CMS:2014fjm}. We consider the leptophilic DM scenario in an effective field theory (EFT) framework~\cite{Brod:2017bsw}. The DM-electron interactions are parametrized by high-dimensional effective operators and the ultraviolet (UV) cutoff scale. The scattering production of a fermionic DM pair will be calculated for effective DM operators of different bilinear products. We will show the sensitivities of laser induced Compton scattering to the effective cutoff scale as a function of the DM mass and compare them with direct detection experiments.

This paper is organized as follows.
In Sec.~\ref{sec:DMEFT}, we review the EFT operators for DM-electron interaction and the characteristics of laser induced Compton scattering for DM pair production.
We present the detailed calculations of the relevant Compton scattering processes for different EFT operators in Sec.~\ref{sec:Calc}. The results of sensitivity reach for the UV cutoff scale are then shown in Sec.~\ref{sec:results}. Our conclusions are drawn in Sec.~\ref{sec:Con}.

\section{DM-electron EFT and laser induced scattering}
\label{sec:DMEFT}

In this section, based on the DM-electron EFT, we consider the method of laser induced Compton scattering and show the characteristics of the scattering producing Dirac-type DM particle pairs in an intense laser field
\begin{eqnarray}
e^-(p) + n\omega (k) \to e^-(p') + \chi (p_\chi) + \overline{\chi} (p'_\chi)\;,
\end{eqnarray}
where $n$ denotes the number of optical photons with energy $\omega$.
In principle, this process can be initiated by some mediator
which connects the DM pair and the electron.
In such a case, the production rates are dominantly determined by the mass
of the mediator due to resonant generation.
The experimental constraints are then given with respect to a combination of the mass parameter
and the couplings between the mediator and the electron as well as the DM.
In this work, we do not aim to perform an analysis based on a UV complete model.
In contrast, for simplicity, we assume that the mediator is heavy enough such that it can be integrated out. The interaction between the electron and the fermionic DM can be described by
low-energy effective operators. Below we review such an effective field theory of DM and
study its phenomenology in terms of the nonlinear Compton scattering.

\subsection{DM-electron effective operators}

The interaction between DM $\chi$ and ordinary matter is described by the effective Lagrangian
\begin{eqnarray}
\mathcal{L}_\chi=\sum_i {1\over \varLambda_i^{d_i-4}} \mathcal{O}_i\;,
\end{eqnarray}
where $d_i$ is the mass dimension of the effective operator $\mathcal{O}_i$.
We take into account the following EFT operators for Dirac-type fermionic DM particles interacting with either a photon or an electron-positron pair~\cite{Brod:2017bsw}
\begin{eqnarray}
&&\mathcal{O}_{MD}=(\overline{\chi}\sigma^{\mu\nu}\chi)F_{\mu\nu}\;,~~~\mathcal{O}_{ED}=(\overline{\chi}\sigma^{\mu\nu}i\gamma_5\chi)F_{\mu\nu}\;,
\label{eq:dipole}
\\
&&\mathcal{O}_{SS}=(\overline{e}e)~ (\overline{\chi}\chi)\;,~~~\mathcal{O}_{SP}=(\overline{e}e)~(\overline{\chi}i\gamma_5\chi)\;,
\label{eq:S}\\
&&\mathcal{O}_{PS}=(\overline{e}i\gamma_5 e)~ (\overline{\chi}\chi)\;,~~~\mathcal{O}_{PP}=(\overline{e}i\gamma_5 e)~ (\overline{\chi}i\gamma_5\chi)\;,
\label{eq:P}\\
&&\mathcal{O}_{VV}=(\overline{e}\gamma^\mu e)~ (\overline{\chi}\gamma_\mu\chi)\;,~~~\mathcal{O}_{VA}=(\overline{e}\gamma^\mu e)~(\overline{\chi}\gamma_\mu\gamma_5\chi)\;,
\label{eq:V}\\
&&\mathcal{O}_{AV}=(\overline{e}\gamma^\mu \gamma_5 e)~ (\overline{\chi}\gamma_\mu\chi)\;,~~~\mathcal{O}_{AA}=(\overline{e}\gamma^\mu \gamma_5 e)~(\overline{\chi}\gamma_\mu\gamma_5\chi)\;.
\label{eq:A}
\end{eqnarray}
The magnetic and electric dipole operators $\mathcal{O}_{MD}$ and $\mathcal{O}_{ED}$ are dimension-5 operators, with $F_{\mu\nu}$ being the electromagnetic field strength tensor. Other dimension-6 operators are formed by either the product of scalar and pseudo-scalar currents, or vector and axial-vector currents. We omit the tensor operators which cannot be generated by tree-level UV models.

The validity of the above effective operators holds as long as
the momentum transfer or the center-of-mass (c.m.) energy is much smaller than the energy scales
$\varLambda_i$. In our case, the c.m. energy is always less than the electron mass.
Hence, the EFT description is valid if the energy scales are in the range of $\varLambda_i\gg 1~{\rm MeV}$.
Furthermore,
the scattering can happen only for DM particles with mass below the electron mass, i.e., $m_\chi \lesssim m_e$. It thus provides a complementary strategy for light DM detection to direct detection experiments~\cite{Goodman:1984dc}. Moreover, in terms of mono-$X$ searches, the validity of
the EFT approach is a longstanding problem at high-energy hadron colliders~\cite{Busoni:2013lha}. At $e^+e^-$ colliders, the obtained lower limit on the UV energy scale is larger than their c.m. energy but not large enough to ensure the EFT validity~\cite{Barman:2021hhg,Kundu:2021cmo}. We will not compare our results with the collider limits.

\subsection{The electron’s Volkov state wave function}

In the presence of an electromagnetic potential, the wave function of a relativistic fermion of mass $m$ is governed by the following Dirac equation
\bee
(i \slashed{\partial} - Qe \slashed{A} -m) \psi(x)=0 \;,
\ene
where $e$ is the unit of electric charge and $Q$ is the charge operator (e.g. $Q\psi=-\psi$ for an electron).
We assume that the electromagnetic potential $A^\mu(x)$ of the incoming laser field
moves along the direction specified by the wave vector $\boldsymbol{k}$
which satisfies the on-shell condition $k^2=0$.
To be specific, the laser wave is assumed to be circularly polarized and monochromatic~\cite{King:2018qbq}. In the Lorentz gauge $k \cdot A=0$, the vector potential $A^\mu$ can be expressed as
\begin{equation}
A^\mu(x)=a\left(\varepsilon_1^\mu \cos\varphi +\varepsilon_2^\mu \sin\varphi\right) \,,
\end{equation}
where the phase $\varphi$ is defined as
$\varphi \equiv k \cdot x = \omega t-\boldsymbol{k} \cdot \boldsymbol{x}$
with $\omega$ being the frequency of the incoming laser,
$\varepsilon_1$ and $\varepsilon_2$ are the two mutually orthogonal polarization vectors,
and $a$ is the amplitude of the laser field which is related to the strength of the laser beam
by the power density
\begin{equation}
I= \frac{1}{4 \pi} a^2 \omega^2 \,.
\end{equation}
The above model of either very short or highly focused light pulses may be oversimplified,
but the essential properties of the laser field are properly taken into account.
In this work, we will use this simplified model to study the laser induced Compton scattering and the production of a DM pair. In a realistic experimental setup, the high-intensity laser would emit the photons in pulses. One can introduce a function $f(x)$ for the pulse shape describing the spatial dependence of the vector potential in the parameterization of $A^\mu$~\cite{Dillon:2018ypt}.

Furthermore, since the laser field is taken as a classical external potential,
the higher-order (nonlinear) effects of the electron decay in the laser field have to be included.
This can be addressed automatically by employing the Volkov state~\cite{Wolkow1935}
which is the exact solution of the Dirac equation in the presence of a circularly polarized laser field.
For a Dirac particle with momentum $p$, the Volkov wave function is given as
\begin{equation}
\psi_{p}
=
\left[1- \frac{e}{2 k \cdot p} \slashed{k} \slashed{A}(\varphi) \right]
u(p) \mathrm{e}^{- \mathrm{i} \varPhi(\varphi) - \mathrm{i} p \cdot x} \,,
\end{equation}
where $u(p) $ is the usual Dirac wave function for a free particle, and
$\varPhi(\varphi)$ is a phase factor depending on $\varphi$
\begin{equation}
\varPhi(\varphi)
=
\int_0^{\varphi=k \cdot x} \mathrm{~d} \varphi'
\left( - \frac{e A(\varphi') \cdot p}{k \cdot p} - \frac{e^2 A^2(\varphi')}{2 k \cdot p}\right) \,.
\end{equation}
On the other hand, due to the direct coupling between the DM and photon in the dipole effective operators given in Eq.~(\ref{eq:dipole}),
the intense laser can potentially induce nonlinear effects on the DM field. As a result, the non-perturbative solution of the DM in the laser field has to be solved.
However, such effects are scaled by the factor $ea/\varLambda$.
We will show that the scale factor is considerably small, $ea/\varLambda \sim 10^{-7}$. Hence, the non-perturbative effects in the dipole operators can be safely neglected.

\subsection{Formalism of laser-induced scattering}

One of the essential properties of the laser induced processes is the position dependence.
In this case, it is more convenient to calculate the scattering matrix in the position representation.
Since the potential is assumed to depend on the space-time coordinates $x$
only through the scalar product $\varphi=k \cdot x$, i.e., $A^\mu(\varphi)=A^\mu(k \cdot x)$,
the scattering matrix element can be written as
\begin{equation}
S_{f i}
=
\int \mathrm{d}^4 x \,
\mathrm{e}^{-\mathrm{i}\left(q   - q' - p_{\chi} - p'_{\chi} \right) \cdot x}
\calm(\varphi) \, \mathrm{e}^{ \mathrm{i} \widetilde{\varPhi}'(\varphi)} \,,
\end{equation}
where $q = p + \frac{e^2 a^2}{2 k \cdot p} k $ and $q' = p'+\frac{e^2 a^2}{2 k \cdot p'} k$
are the effective momenta of the incoming and outgoing electrons in the laser field, respectively, $p_{\chi}$ and $p'_{\chi}$ are the momenta of the outgoing DM particle and its anti-particle,
respectively, $\calm(\varphi)$ is the corresponding amplitude,
and $\mathrm{e}^{ i \widetilde{\varPhi}'}(\varphi)$ is the possible phase factor.
After employing the Volkov wave functions for both the incoming and outgoing electrons, for dimension-6 operators,
one can easily find that the amplitude is given as
\begin{equation}
\calm
=
\frac{1}{ \varLambda_{XX'}^2 }
\Bigg[ \overline{u(p')}
\left(1-\frac{e \slashed{A} \slashed{k} }{2 k \cdot p^{\prime}}\right)
\varGamma_{X}
\left(1- \frac{e \slashed{k} \slashed{A} }{2 k \cdot p}\right) u(p)
\Bigg] \Bigg[
\overline{u_\chi(p_{\chi})} \varGamma_{X'} v(p'_{\chi})
\Bigg] \,,
\end{equation}
where $\varGamma_{X}$ and $\varGamma_{X'}$ are the possible Lorentz structures
of the effective operators.
For dimension-5 operators, the amplitude reads as
\begin{eqnarray}
\calm
={-2(p_\chi+p'_\chi)_\nu\over (p_\chi+p'_\chi)^2} \frac{1}{ \varLambda_{X'} }
\Bigg[ \overline{u(p')}
\left(1-\frac{e \slashed{A} \slashed{k} }{2 k \cdot p^{\prime}}\right)
e\gamma_\mu
\left(1- \frac{e \slashed{k} \slashed{A} }{2 k \cdot p}\right) u(p)
\Bigg] \Bigg[
\overline{u_\chi(p_{\chi})} \varGamma_{X'} v(p'_{\chi})
\Bigg] \,,
\end{eqnarray}
where $\varGamma_{X'}=\sigma^{\mu\nu}$ and $\varLambda_{X'}=\varLambda_{MD}$ or $\varLambda_{ED}$.
The corresponding phase is given as
\begin{equation}
\widetilde{\varPhi}'
=
e a\left(\frac{\varepsilon_1 \cdot p}{k \cdot p}-\frac{\varepsilon_1 \cdot p^{\prime}}{k \cdot p^{\prime}}\right) \sin \varphi
-
e a\left(\frac{\varepsilon_2 \cdot p}{k \cdot p}-\frac{\varepsilon_2 \cdot p^{\prime}}{k \cdot p^{\prime}}\right) \cos \varphi\;.
\end{equation}
We then define an ``effective momentum'' for the incoming electron as
\begin{equation}
q^\mu=p^\mu+\frac{e^2 a^2}{2 k \cdot p} k^\mu\;.
\end{equation}
Similarly, the momenta of the final electron in the laser field can be obtained with the substitution $p\to p'$ and $q\to q'$.
By introducing the following transverse momentum ($Q\cdot k = 0$)
\begin{equation}
Q=\frac{q'}{k \cdot q'}-\frac{q}{k \cdot q}\;,
\end{equation}
the phase can be written as
\begin{equation}
\widetilde{\varPhi}'
= e a\left( - \varepsilon_1 \cdot Q \sin \varphi + \varepsilon_2 \cdot Q \cos \varphi\right)
= - z \sin \left(\varphi-\varphi_0\right)\;,
\end{equation}
where
\bea
z
&=&
e a \sqrt{\left(\varepsilon_1 \cdot Q\right)^2+\left(\varepsilon_2 \cdot Q\right)^2}
= e a \sqrt{-Q^2} \,,
\\[3mm]
\cos \varphi_0
&=&
e a \frac{\varepsilon_1 \cdot Q}{z} \,,
\\[3mm]
\sin \varphi_0
&=&
e a \frac{\varepsilon_2 \cdot Q}{z} \,.
\ena
The phase factor $\mathrm{e}^{\mathrm{i} \widetilde{\varPhi}' }$
is a periodic function of the variable $\varphi$
and thus can be expanded into a discrete Fourier series.
We have the following ansatz
\begin{equation}
\mathrm{e}^{ - \mathrm{i} z \sin \left(\varphi-\varphi_0\right)}
=
\sum_{n=-\infty}^{\infty} B_n(z) \mathrm{e}^{-\mathrm{i} n \varphi}
\end{equation}
where $B_n(z) = J_n(z) \mathrm{e}^{\mathrm{i} n \varphi_0}$ with
$J_n(z)$ being the Bessel function.
With the help of the above transformations, one can easily find that
the $S$-matrix element can be expanded as
\begin{equation}
S_{f i}
=
\sum_{n=-\infty}^{\infty} \int \mathrm{d}^4 x
~\mathrm{e}^{-\mathrm{i}\left(q + nk - q^{\prime} - p_{\chi} - p'_{\chi} \right) \cdot x}
\calm_n(z) \,,
\end{equation}
with the $n$-th amplitude given as
\bee
\calm_n(z) = B_n(z) \calm(z) \,.
\ene
Usually, the coordinate dependence of the amplitude $\calm(z)$,
and hence the $n$-th amplitude $\calm_n(z)$, can be eliminated.
As a result, the space-time integration now simply reduces to
\begin{equation}
\int \mathrm{d}^4 x~\mathrm{e}^{-\mathrm{i}
\left(q + nk - q^{\prime} - p_{\chi} - p'_{\chi} \right) \cdot x}
=
(2 \pi)^4 \delta^4\left(q + nk - q^{\prime} - p_{\chi} - p'_{\chi} \right) \,.
\label{eq:4mom}
\end{equation}
Clearly, the summation variable $n$ can be viewed as the (net) number of laser photons that are absorbed $(n>0)$ or emitted $(n<0)$ in the process. This is an interesting result because the electromagnetic wave was originally introduced as a classical external field. The discretization of four-momentum evident in Eq.~(\ref{eq:4mom}) arises from the periodicity of the plane wave in space and time.

Since the kinematics of the outgoing DM fermions, $\chi$ and $\bar\chi$,
can never be measured in practice,
their relative motion should be integrated out.
This can be done by introducing a fictitious momentum $k' = p_{\chi} + p'_{\chi}$.
The corresponding 3-body phase space is decomposed as follows
\bee
d \varPi_3
=
\int \frac{ d m_{k'}^2 }{2\pi}
d \varPi_2(q + nk - q^{\prime} - k')
d \varPi_2(k' - p_{\chi} - p'_{\chi}) \,,
\ene
where $m_{k'}^2 = k'^2 \ge 4m_\chi^2$, and $\varPi_2$ is the standard 2-body phase space. Here and hereafter we will use $d\varPi_{P, n}$ and $d\varPi_{D,n}$ to denote the phase space of
the production $d \varPi_2(q + nk - q^{\prime} - k')$ and the
decay $d \varPi_2(k' - p_{\chi} - p'_{\chi})$, respectively.
For convenience, we introduce following Lorentz invariant
parameter to simplify the above 2-body phase space
\bee
u = \frac{k \cdot q}{ 2 k \cdot q' }  \,.
\ene
In terms of the parameter $u$, the production 2-body phase space is given by
\bee
d\varPi_{P,n} = \frac{1}{32\pi^2 u^2} du d\phi^\ast_n \,,
\ene
with $\phi^\ast_n$ being the azimuthal angle of the momentum $k'$.
Further details of our parameterization of the production phase space
can be found in Ref.~\cite{Ma:2024ywm}.
The amplitude can be calculated in the rest frames of $q + nk$ and $k'$, respectively.
Since it involves different Lorentz structures, we study them separately in the following section.

\section{Calculation of laser induced Compton scattering to DM pair}
\label{sec:Calc}

In this section, we calculate in detail the laser induced Compton scattering to DM pair. Our calculations are categorized based on the effective operators of different bilinear products. For convenience, we adopt the method of helicity amplitude in terms of the density matrices of production and decay.

\subsection{scalar and pseudo-scalar bilinears}

In cases of the operators $\mathcal{O}_{SS}$, $\mathcal{O}_{SP}$, $\mathcal{O}_{PS}$ and $\mathcal{O}_{PP}$ made of scalar and pseudo-scalar bilinears ($X, X'=S, P$),
the amplitude can be decomposed as follows
\bee
\calm_{XX'} = \frac{1}{ \varLambda_{XX'}^2 } \calm^P_{X} \calm^D_{X'} \,,
\ene
with
\bea
\calm^P_{X}
& = &
\Bigg[ \overline{u(p')}
\left(1-\frac{e \slashed{A} \slashed{k} }{2 k \cdot p^{\prime}}\right)
\varGamma_{X}
\left(1- \frac{e \slashed{k} \slashed{A} }{2 k \cdot p}\right) u(p)
\Bigg] \,,
\\[3mm]
\calm^D_{X'}
&=&
\Bigg[
\overline{u_\chi(p_{\chi})} \varGamma_{X'} v(p'_{\chi})
\Bigg] \,.
\ena
Here the amplitudes $\calm^P_{X}$ and $\calm^D_{X'}$ describe the production and decay
of a fictitious spin-0 particle with momentum $k'$, respectively,
and both of them are Lorentz invariant.
Since only the production amplitude $\calm^P_{X}$ depends on the variable $z$,
the $n$-th amplitude can be written as
\bee
\calm_{X X', n} = \frac{1}{ \varLambda_{XX'}^2 } \calm^{P}_{X, n} \calm^D_{X'} \,,
\ene
where $\calm^{P}_{X, n} = B_n\calm^P_{X}$.
It should be noted that the amplitude $\calm^D_{X'}$ is also $n$-dependent
as the outgoing momenta $p_\chi$ and $p'_\chi$ are subject to
the absorption strength of the incoming laser.
This becomes clear when we interpret the $n$-th amplitude as corresponding to absorbing
$n$ photons from the laser field. However, here and hereafter, the $n$-dependence of the decay amplitude $\calm^D_{X'}$ is always suppressed,
unless its explicit expression is shown.

The total density matrix elements are given as
\bee
\rho_{XX', n} = \frac{1}{ \varLambda_{XX'}^4 }  \calp_{X, n} \cald_{X'} \,,
\ene
where
\bea
\calp_{X, n}
&=&
\overline{ \big| \calm^{P}_{X, n} \big|^2  }
= \frac{1}{2} \calm^{P}_{X, n} \left( \calm^{P}_{X, n} \right)^\dag  \,,
\\[3mm]
\cald_{X'}
&=&
\overline{ \big|  \calm^D_{X'}  \big|^2 }
= \sum_{\lambda_{\chi}, \lambda_{\bar\chi}}
\calm^D_{X'}  \left( \calm^D_{X'} \right)^\dag \,.
\ena
Here the spin of the incoming electron is averaged,
and the spin of the outgoing dark fermions has been summed over.
The total decay width is then given as
\bee
\varGamma_{XX'}
=
\frac{1}{ \varLambda_{XX'}^4 }
\sum_{n=-\infty}^{\infty}
\frac{1}{2 Q_{\rm Lab}} \int d \varPi_{P,n}
\int \frac{ d m_{k'}^2 }{2\pi}
\calp_{X, n}  \; \overline{ \cald_{X',n}  } \,,
\ene
where $Q_{\rm Lab}$ is the energy of the incoming dressed electron in the laboratory frame, and $\overline{ \cald_{X'}  } = \int d \varPi_{D, n}  \cald_{X'} $
is the averaged decay density matrix
over the phase space.
For the production density matrix,
after tedious but straightforward calculations, we arrive at the following results
\bea
\calp_{S, n}
&=&
4 \left( m_e^2 - \frac{1}{4} m_{k'}^2 \right) J_n^2
-
2e^2 a^2 u \left( 1- \frac{1}{ 2 u } \right)^2
\calj_n  \,,
\\[3mm]
\calp_{P, n}
&=&
-m_{k'}^2  J_n^2
-
2 e^2 a^2 u \left(1 - \frac{1}{ 2u } \right)^2
\calj_n \,,
\ena
with
\bee
\calj_n = J_n^2  - \dfrac{1}{2}\Big(J_{n-1}^2+J_{n+1}^2  \Big) \,.
\ene
In either case, there are terms explicitly proportional to $a^2$ for pure nonlinear absorption of the laser field.
One can also see that there is an additional contribution
proportional to $m_e^2J_n^2$ in the scalar production density matrix.
For both scalar and pseudo-scalar operators,
the non-trivial amplitudes are given by the spinors with chiral-flipping combinations, i.e., ``RL'' and ``LR'',
whose amplitudes are proportional to the electron mass.
However, while the chiral-flipping amplitude for the scalar operator adds constructively,
the one for the pseudo-scalar operator adds deconstructively.
As a result, the scalar production density matrix receives an additional contribution proportional to $m_e^2$.
The decay density matrices can be easily calculated.
After integrating out the 2-body phase space of the outgoing DM fermions $\chi$ and $\overline{\chi}$,
the averaged decay density matrices are given by
\bea
\overline{\cald_{S}}
&=&
\frac{1}{4\pi} m_{k'}^2 \beta_\chi^{3} \,,
\\[3mm]
\overline{\cald_{P}}
&=&
\frac{1}{4\pi} m_{k'}^2 \beta_\chi \,,
\ena
where $\beta_\chi = \sqrt{1 - 4m_\chi^2/m_{k'}^2 }$ with $m_{k'} \ge 2 m_\chi$ is
the velocity of the outgoing fermionic DM.
Similar to the production case, the decay amplitudes are also constructively and destructively
added for the scalar and pseudo-scalar operator, respectively.
As a result, the scalar operator can induce only $P$-wave decay,
in contrast the leading order nontrivial decay amplitude for the pseudo-scalar operator is given by the $S$-wave.
Consequently, the averaged decay density matrices have difference power dependencies
on the velocity $\beta_\chi$.

Because of the non-trivial dependence on the invariant mass $m_{k'}$,
it is difficult to obtain the analytical expressions of the total decay width.
Here we discuss their physical properties with the help of numerical integration.
We take the same experimental setup as the Laser Und XFEL Experiment (LUXE)~\cite{LUXE:2023crk},
in which the incoming electron beam has an energy of $E_{\rm Lab}=14$ GeV
and a laser beam of green light with $\omega_{\rm Lab}=2.35$ eV
as a benchmark to study the physical properties~\footnote{Note that we omit the small scattering angle $\theta_{\rm Lab}=17^\circ$ and assume the two beams are head-on colliding in our calculation.}.

\begin{figure}[thb!]
\begin{center}
\includegraphics[width=0.8\textwidth]{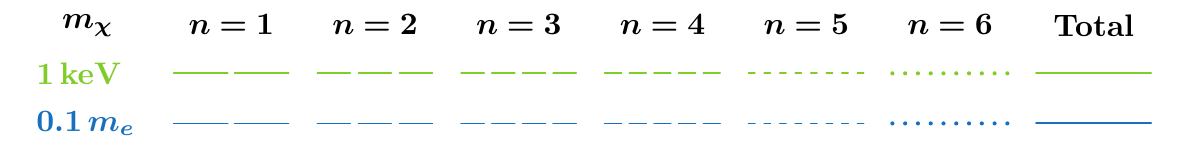}
\\
\includegraphics[width=0.48\textwidth]{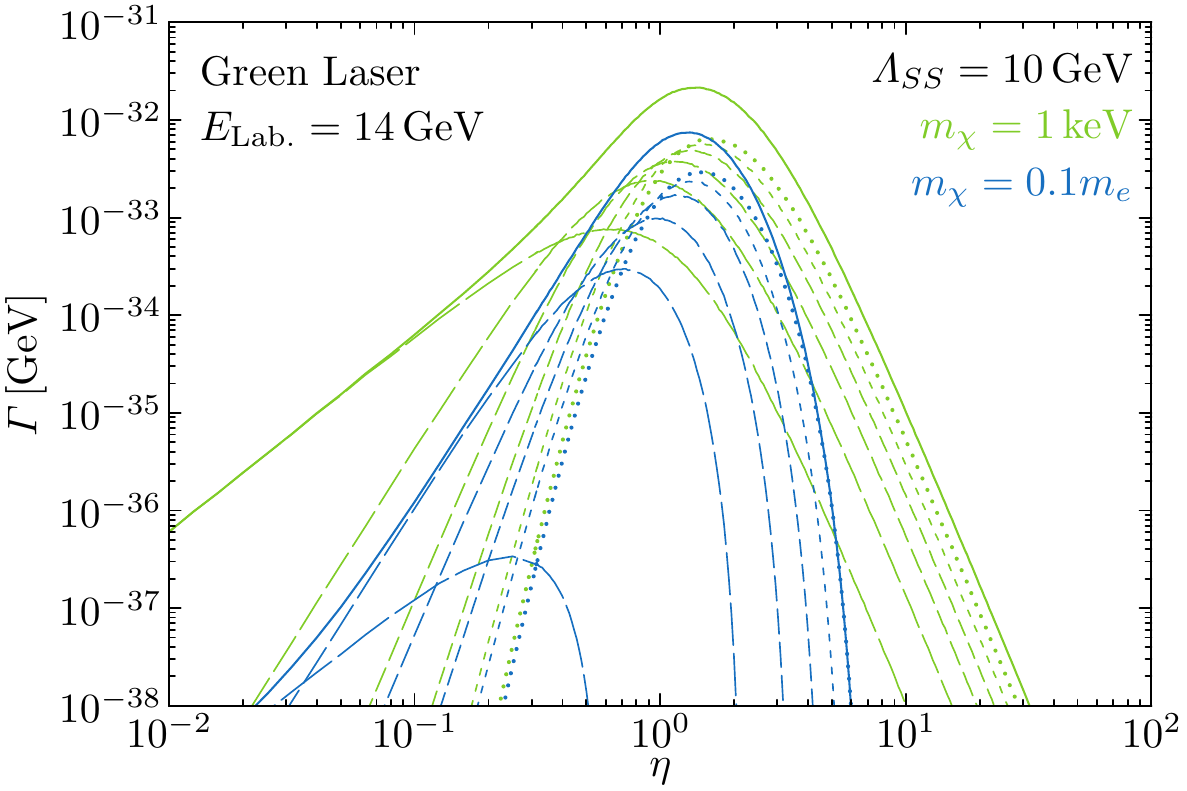}
\includegraphics[width=0.48\textwidth]{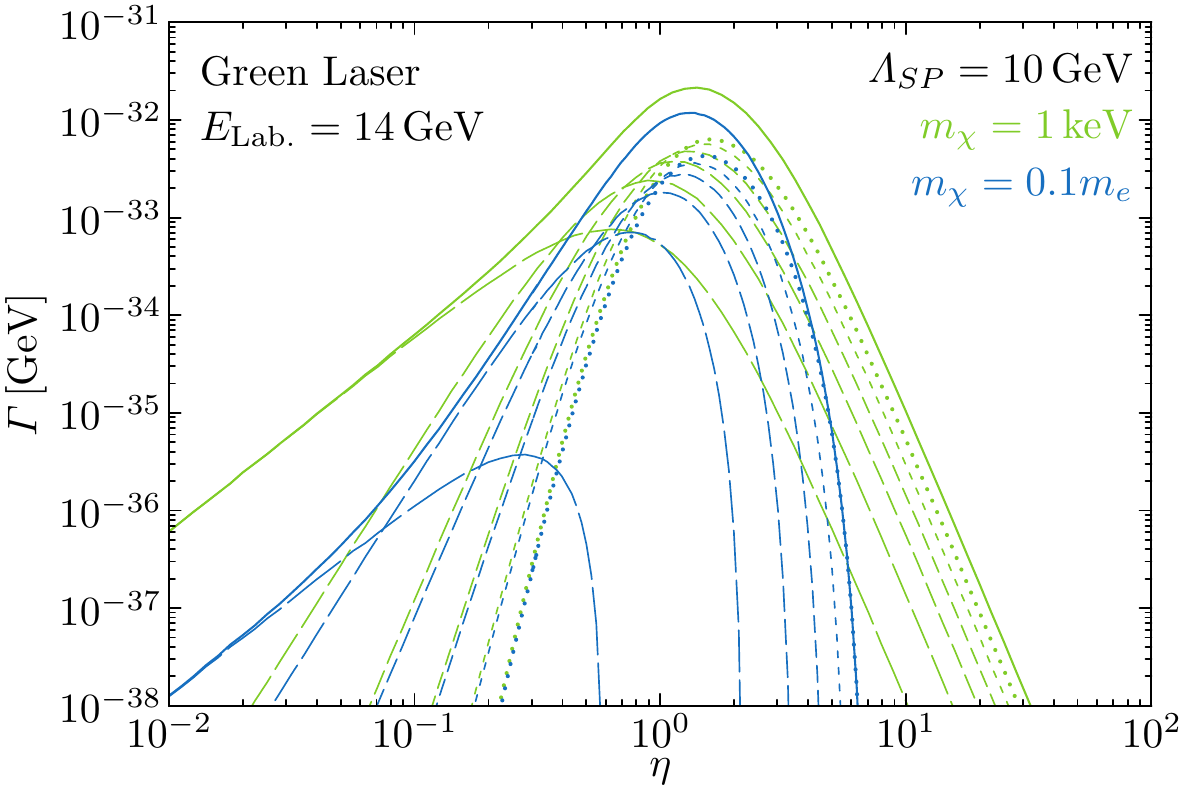}
\\
\includegraphics[width=0.48\textwidth]{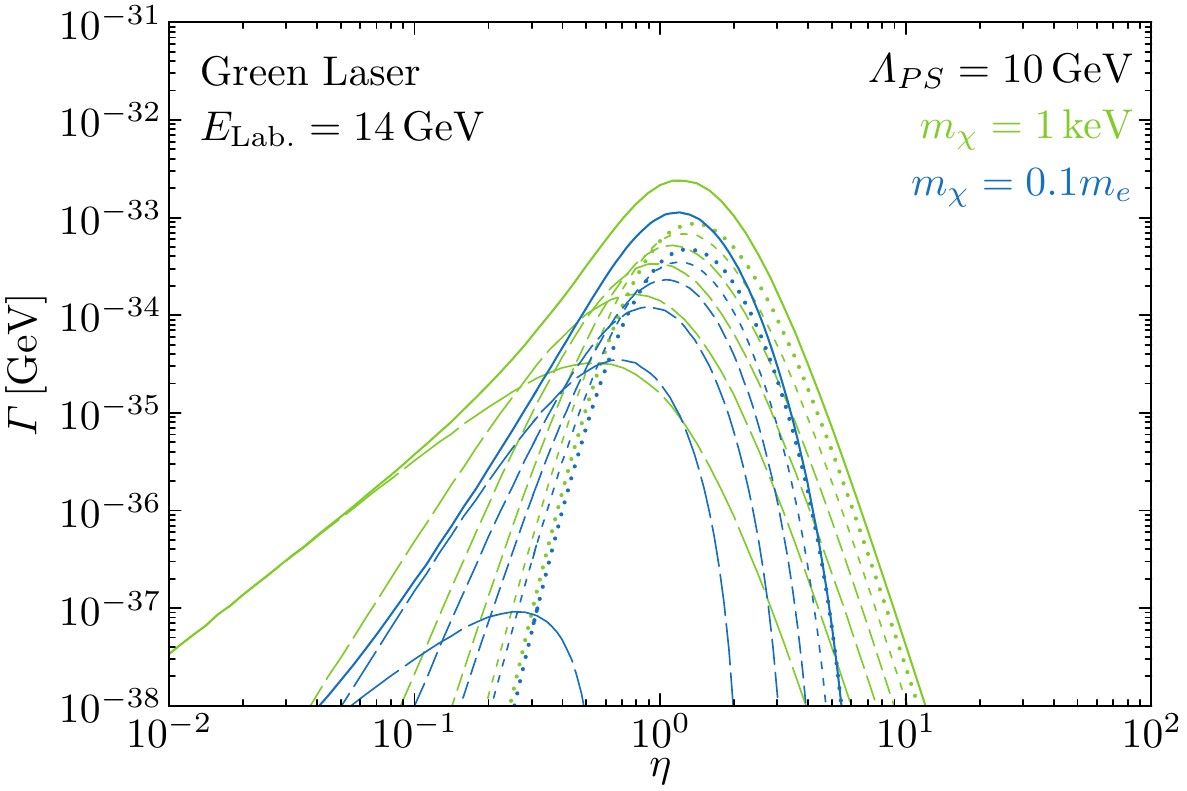}
\includegraphics[width=0.48\textwidth]{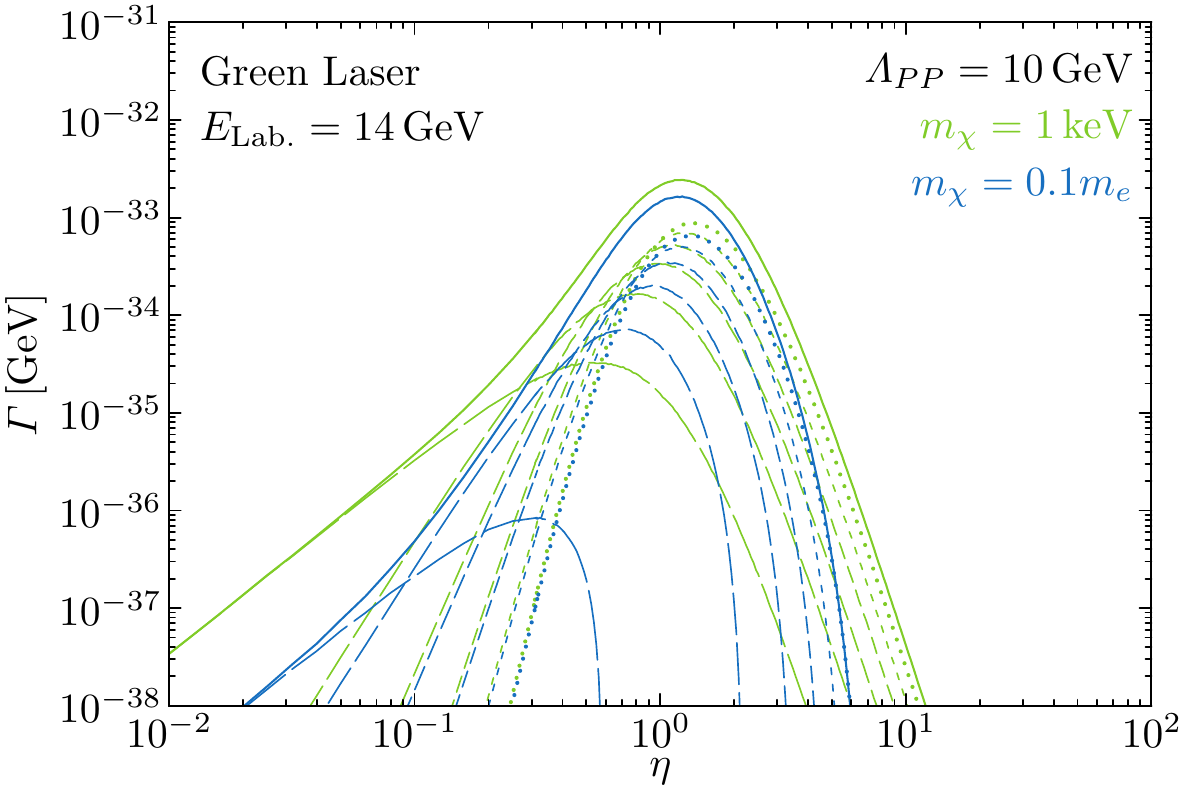}
\caption{
The total decay widths of the laser dressed electron for the operator $\calo_{SS}$ (top-left),
$\calo_{SP}$ (top-right), $\calo_{PS}$ (bottom-left) and $\calo_{PP}$ (bottom-right)
as a function of laser intensity parameter $\eta$.
The results are shown for a fiducial energy scalar $\varLambda = 10$ GeV,
and a DM mass $m_\chi = 1$ keV (green curves) and $m_\chi = 0.1m_e$ (blue curves).
The total contributions are shown by solid curves,
while the contributions up to $n=6$ are also shown by non-solid curves. See the legend on top for the label details.
}
\label{fig:SP:Eta}
\end{center}
\end{figure}

Fig.~\ref{fig:SP:Eta} shows the decay widths of the laser dressed electron
for the operator $\calo_{SS}$ (top-left), $\calo_{SP}$ (top-right), $\calo_{PS}$ (bottom-left)
and $\calo_{PP}$ (bottom-right)
as a function of the laser intensity parameter
\bee
\eta\equiv {ea\over m_e}={e \varepsilon_0\over \omega_{\rm Lab} m_e}\;,
\ene
where $\omega_{\rm Lab}$ denotes the laser beam energy in the laboratory frame.
The results are shown for a fiducial energy scale $\varLambda = 10$ GeV, and a DM mass $m_\chi = 1$ keV (green curves) or $m_\chi = 0.1m_e$ (blue curves).
The total contributions are shown by solid curves,
and the contributions from the $n$-th branch of laser photon are shown by long-dashed $(n=1)$,
long-dash-dotted $(n=2)$, dashed $(n=3)$, dash-dotted $(n=4)$, big dotted $(n=5)$
and dotted $(n=6)$ curves. One can clearly see that for small values of $\eta$,
the dominant contribution is given by the $n=1$ branch.
The contributions of the higher-order absorption grow quickly as increasing $\eta$,
and reach a maximum around $\eta \sim 2$.
For every branch, there is a maximum for the intensity parameter $\eta$
\bee
\eta^{\rm max}_n
=
\Big[\Big({2n\omega_{\rm Lab}(E_{\rm Lab}+\cos\theta_{\rm Lab}p_{\rm Lab})- 4m_\chi^2\over 4m_e m_\chi}\Big)^2-1\Big]^{1/2}\;,~~~p_{\rm Lab}=\sqrt{E_{\rm Lab}^2-m_e^2}\;,
\ene
beyond which there is no enough phase space for the decay.
This explains both the width reduction at large $\eta$ and the kinematic suppression visible in the plot.
For lighter DM ($m_\chi = 1$ keV), higher-order contributions can enhance the total decay width by over an order of magnitude.
One can also see that, for heavier DM $\chi$ ($m_\chi = 0.1m_e$),
the higher-order contributions for given intensity parameter become more important.

\begin{figure}[htb!]
\begin{center}
\includegraphics[width=0.48\textwidth]{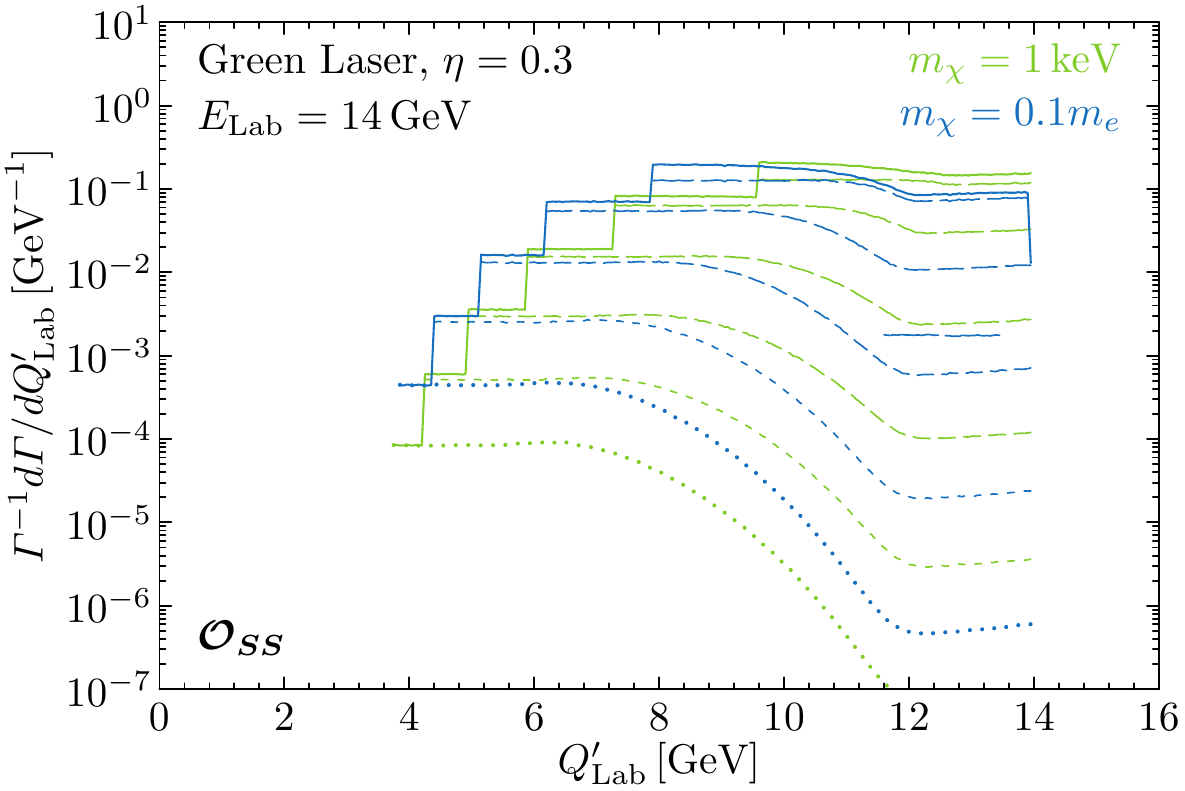}
\includegraphics[width=0.48\textwidth]{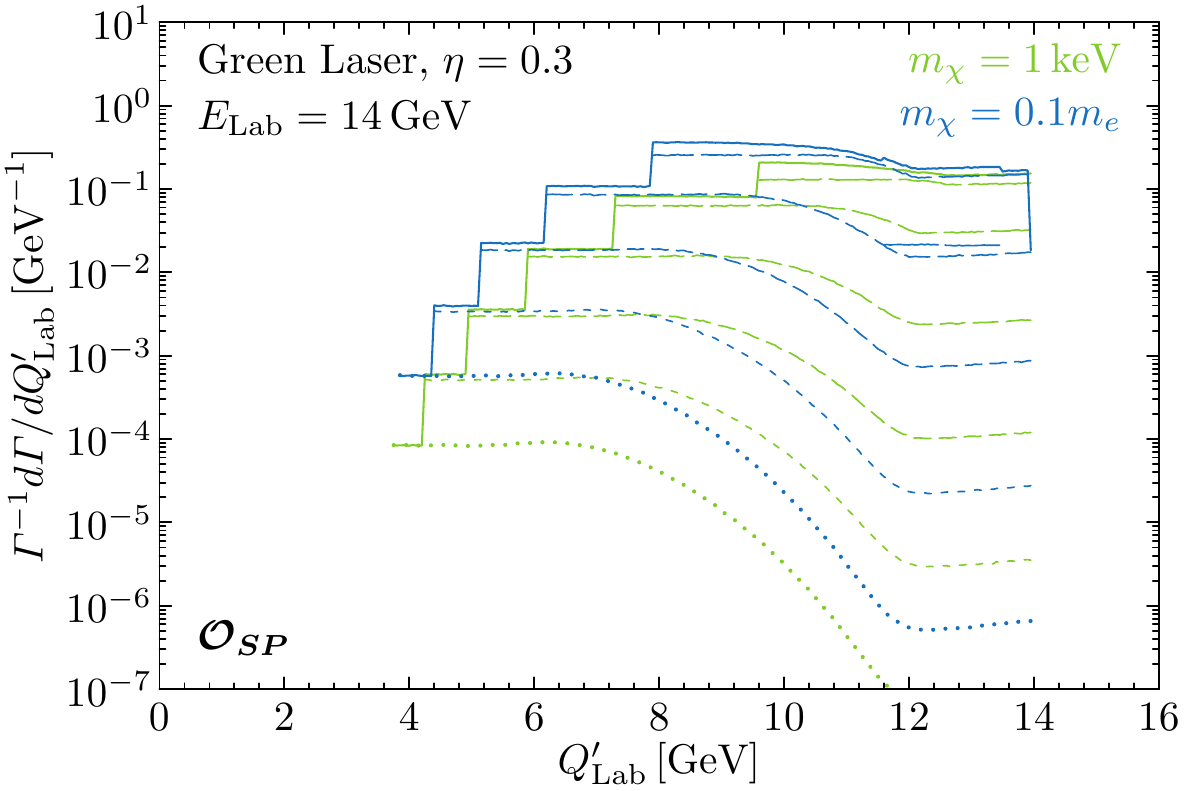}
\\
\includegraphics[width=0.48\textwidth]{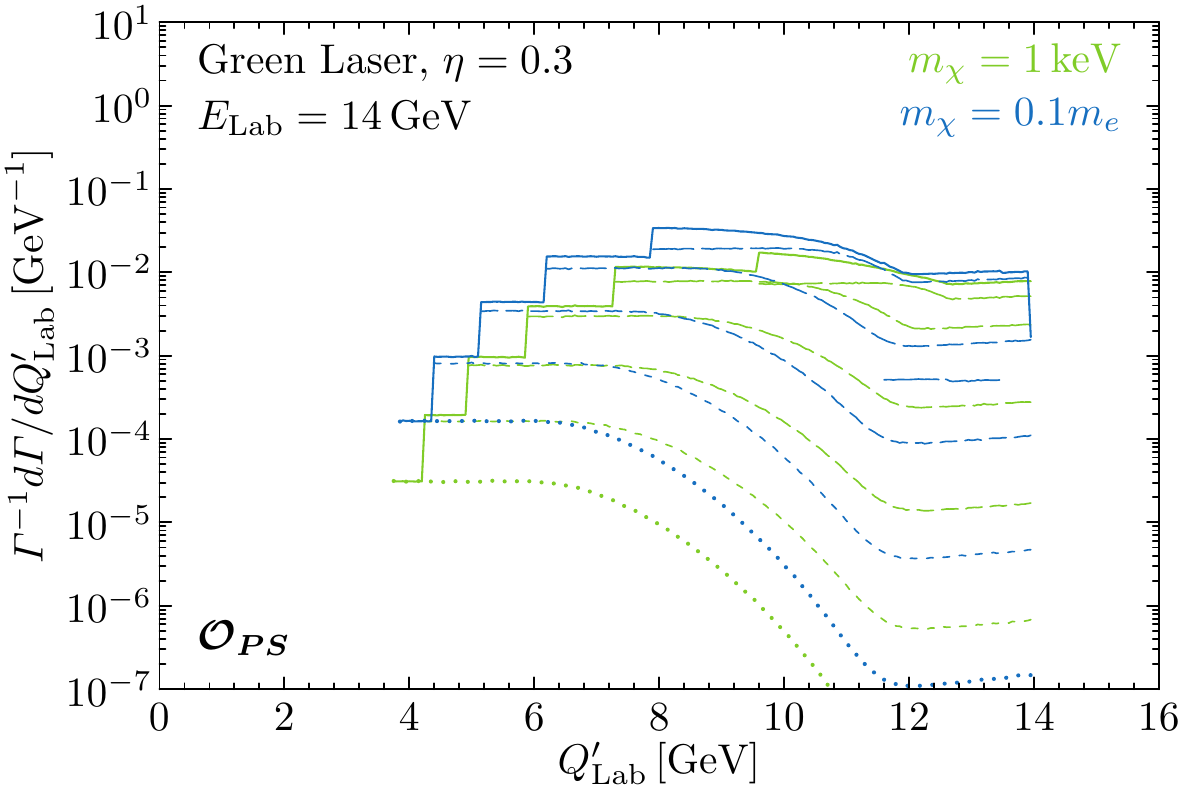}
\includegraphics[width=0.48\textwidth]{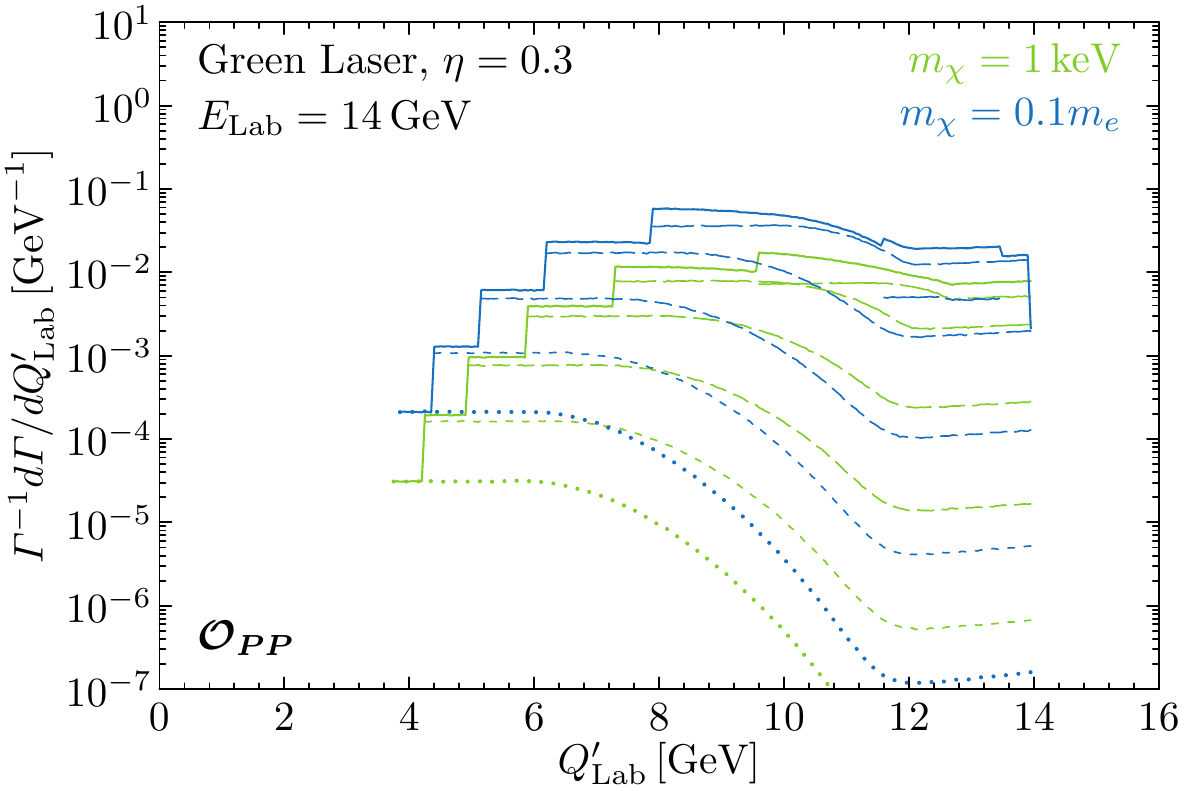}
\caption{The normalized distribution of the decay width with respect to the outgoing electron energy for the operator $\calo_{SS}$ (top-left),
$\calo_{SP}$ (top-right), $\calo_{PS}$ (bottom-left) and $\calo_{PP}$ (bottom-right), as labeled in Fig.~\ref{fig:SP:Eta}.
}
\label{fig:SP:EE}
\end{center}
\end{figure}

Fig.~\ref{fig:SP:EE} shows the normalized distributions of the total decay width
with respect to the outgoing electron energy for the DM masses of $m_\chi = 1$ keV (green curves) and $m_\chi = 0.1m_e$ (blue curves).
The laser intensity parameter is set to the typical value from the SLAC experiment~\cite{Bamber:1999zt},
namely $\eta=0.3$. The total contributions up to $n=6$ are shown by solid curves,
with individual branch contributions shown by non-solid curves.
From Fig.~\ref{fig:SP:EE}, one can see that even at this typical intensity ($\eta=0.3$),
higher-order absorption channels can also give a comparable contribution to the $n=1$ branch in the high-energy region.

One can also find that the higher-order contributions are dominant in the low-energy region.
Even though the laser dressed electron can be at rest in the rest frame of $q+nk$,
it is always moving in the laboratory frame.
Hence, the outgoing electron energy has a minimum in the laboratory frame for each branch,
below which there is not enough phase space to produce the DM pair.
Furthermore, for each branch the contribution decreases with increasing electron energy $Q'_{\rm Lab}$.
The shapes of each branch are different for different operators.

\subsection{vector and axial-vector bilinears}

In cases of the vector and axial-vector operators ($X, X'=V, A$),
the total amplitude can be written as a contraction of two vector currents
\bee
\calm_{XX'} = \frac{1}{ \varLambda_{XX'}^2 } g^{\mu\nu} \calm^P_{X, \mu} \calm^D_{X', \nu} \,,
\ene
with $\calm^P_{X, \mu}$ and $\calm^D_{X, \nu}$ being the currents
for the production and decay of a fictitious spin-1 particle with momentum $k'$,
respectively
\bea
\calm^P_{X, \mu}
& = &
\Bigg[ \overline{u(p')}
\left(1-\frac{e \slashed{A} \slashed{k} }{2 k \cdot p^{\prime}}\right)
\varGamma_{X}
\left(1- \frac{e \slashed{k} \slashed{A} }{2 k \cdot p}\right) u(p)
\Bigg] \,,
\\[3mm]
\calm^D_{X', \nu}
&=&
\Bigg[
\overline{u_\chi(p_{\chi})} \varGamma_{X'} v(p'_{\chi})
\Bigg] \,,
\ena
where $\varGamma_{X(X')} = \gamma^\mu, \gamma^\mu\gamma_5$
stand for the corresponding vector and axial vector Lorentz structures.
Unlike the case of (pseudo-)scalar operator,
the production and decay parts for the (axial-)vector operators
are not Lorentz invariant individually.
However, by employing the helicity amplitude method,
the amplitude can be decomposed into two Lorentz invariant amplitudes
along the momentum $k'$.
This can be addressed by inserting the following relation for the metric tensor
\bee
\label{eq:metric}
g^{\mu\nu}
=
\sum_{\lambda = s, 0, \pm1} \eta_{\lambda}
\varepsilon_{\lambda}^{\mu\ast}\varepsilon^\nu_{\lambda}\;,
\ene
where $\eta_s = 1$ and $\eta_{0,\pm1} = -1$,
and $\varepsilon_\lambda^\mu$ are polarization vectors with helicity
$\lambda = s, 0, \pm1$ projected along the momentum $k'$.
One can easily find that the total amplitude can be rewritten as
\bee
\calm_{XX'} = \frac{1}{ \varLambda_{XX'}^2 }
\sum_{\lambda = s, 0, \pm1} \eta_{\lambda} \calm^P_{X, \lambda} \calm^D_{X', \lambda} \,,
\ene
with $\calm^P_{X, \lambda} = \varepsilon_{\lambda}^{\ast} \cdot \calm^P_{X}$ and
$\calm^D_{X', \lambda} = \varepsilon_{\lambda} \cdot \calm^D_{X'}$.
It is clear that both $\calm^P_{X, \lambda}$ and $\calm^D_{X', \lambda}$
are invariant under Lorentz boost along the direction of the momentum $k'$.
By virtue of this,
the decay amplitude can be calculated in the rest frame of the momentum $k'$.
As we have mentioned, even though both
$\calm^P_{X, \lambda}$ and $\calm^D_{X', \lambda}$
depend on the number of laser photons $n$,
only the production amplitude $\calm^P_{X}$ depends on the variable $z$.
Hence, for each photon number $n$ the above decomposition is still valid.
As a result, for a given laser photon number $n$,
the $n$-th amplitude can be written as
\bee
\calm_{XX', n} = \frac{1}{ \varLambda_{XX'}^2 }
\sum_{\lambda = s, 0, \pm1} \eta_{\lambda}  \calm^{P}_{X, n, \lambda} \calm^D_{X', \lambda} \,,
\ene
with $\calm^{P}_{X, n, \lambda} = B_n\calm^P_{X, \lambda}$.

The corresponding total density matrix elements are given as
\bee
\rho_{XX', n}
=
\frac{1}{ \varLambda_{XX'}^2 } \sum_{\lambda, \lambda' = s, 0, \pm1}
\eta_{\lambda} \eta_{\lambda'}
\calp_{X, n, \lambda\lambda'} \cald_{X', \lambda\lambda'} \,,
\ene
with
\bea
\calp_{X, n, \lambda\lambda' }
&=&
\frac{1}{4}  \calm^{P}_{X, n, \lambda} \left( \calm^{P}_{X, n, \lambda'} \right)^\dag \,,
\\[3mm]
\cald_{X', \lambda\lambda'}
&=&
\calm^D_{X', \lambda} \left( \calm^D_{X', \lambda'} \right)^\dag \,.
\ena
Again, the spin of the incoming electron has been averaged,
and the spins of the outgoing dark fermions have been summed over implicitly.
The total decay width is then given as
\bee
\varGamma_{XX'}
=
\frac{1}{ \varLambda_{XX'}^4 }
\sum_{n=-\infty}^{\infty} \frac{1}{2 Q_{\rm Lab}}
\int d \varPi_{P,n}\int \frac{ d m_{k'}^2 }{2\pi}
\sum_{\lambda, \lambda' = s, 0, \pm1}
\eta_{\lambda} \eta_{\lambda'}
\calp_{X, n, \lambda\lambda' }  \; \overline{ \cald_{X', \lambda\lambda' }  } \,,
\ene
with $\overline{ \cald_{X', \lambda\lambda'}  } = \int d \varPi_{D, n}  \cald_{X', \lambda\lambda'} $
being the averaged decay density matrix
over the 2-body phase space of the outgoing DM particles.
Because of this, polarization effects are washed out and
the averaged density matrix $\overline{ \cald_{X', \lambda\lambda' }  }$
is simplified significantly.
After some calculations, one can easily find that
the averaged decay density matrix elements are given as
\bea
\overline{ \cald_{V, \lambda\lambda' }  }
&=&
\delta_{\lambda\lambda'}
\begin{cases}
0  \,, & \text{for } \lambda= s
\\[3mm]
\dfrac{1}{4\pi} m_{k'}^2 \beta_\chi
\left( 1 - \dfrac{1}{3}\beta^2_\chi \right) \,,
&
\text{for } \lambda= 0, \pm1
\end{cases}
\\[3mm]
\overline{ \cald_{A, \lambda\lambda' }  }
&=&
\delta_{\lambda\lambda'}
\begin{cases}
\dfrac{1}{4\pi} m_{k'}^2 \beta_\chi
\left( 1 - \beta^2_\chi \right) \,,  & \text{for } \lambda= s
\\[5mm]
\dfrac{1}{6\pi} m_{k'}^2 \beta_\chi^3 \,,
&
\text{for } \lambda= 0, \pm1
\end{cases}
\ena
One can see that, for either the vector or axial-vector operator,
the averaged decay density matrices with scalar and vector polarization states
have completely different dependence on the velocity
due to different partial wave contributions.
Furthermore, the matrices for the vector state are always diagonal, i.e., proportional to $\delta_{\lambda\lambda'}$,
due to the fact that interference effects have been integrated out.
With the help of this simplification, the total decay width is given as
\bee
\varGamma_{XX'}
=
\frac{1}{ \varLambda_{XX'}^4 }
\sum_{n=-\infty}^{\infty} \frac{1}{2 Q_{\rm Lab}}
\int d \varPi_{P,n}\int \frac{ d m_{k'}^2 }{2\pi}
\sum_{\lambda= s, 0, \pm1}
\calp_{X, n, \lambda\lambda}  \; \overline{ \cald_{X', \lambda\lambda}  } \,,
\ene
which involves only diagonal elements of the production density matrix.
After tedious but straightforward calculations, we arrive at the following results
\bea
\label{eq:prod:v}
\calp_{V, n, ss }
&=&
0 \,,
\\[3mm]
\sum_{\lambda= 0, \pm1}
\calp_{V, n, \lambda\lambda}
&=&
- 4J_n^2 ( m_e^2 + \dfrac{1}{2} m_{k'}^2 )
- 4 e^2 a^2 \calj_n
\left( u + \dfrac{1}{4u} \right) \,,
\\[3mm]
\calp_{A, n, ss }
&=&
- 4 J_n^2 m_e^2
- 8 e^2 a^2 \dfrac{ m_e^2 }{m_{k'}^2}
\calj_n u \left( 1 - \dfrac{1}{ 2u } \right)^2  \,,
\\[3mm]
\sum_{\lambda= 0, \pm1}
\calp_{A, n, \lambda\lambda}
&=&
8 J_n^2 ( m_e^2 - \dfrac{1}{4} m_{k'}^2 )
-
8 e^2 a^2 \calj_n
\left[ \dfrac{1}{2}\left( u + \dfrac{1}{ 4u } \right) + \dfrac{ m_e^2  }{m_{k'}^2}u \left( 1 - \dfrac{1}{ 2 u } \right)^2  \right]\,.
\ena
One can clearly see that for both the vector and axial-vector operators,
there are terms explicitly proportional to $a^2$
due to pure nonlinear absorption of the laser photons,
and terms proportional to $m_e^2J_n^2$ which account for higher-order
absorption of the laser photons.

\begin{figure}[htb!]
\begin{center}
\includegraphics[width=0.48\textwidth]{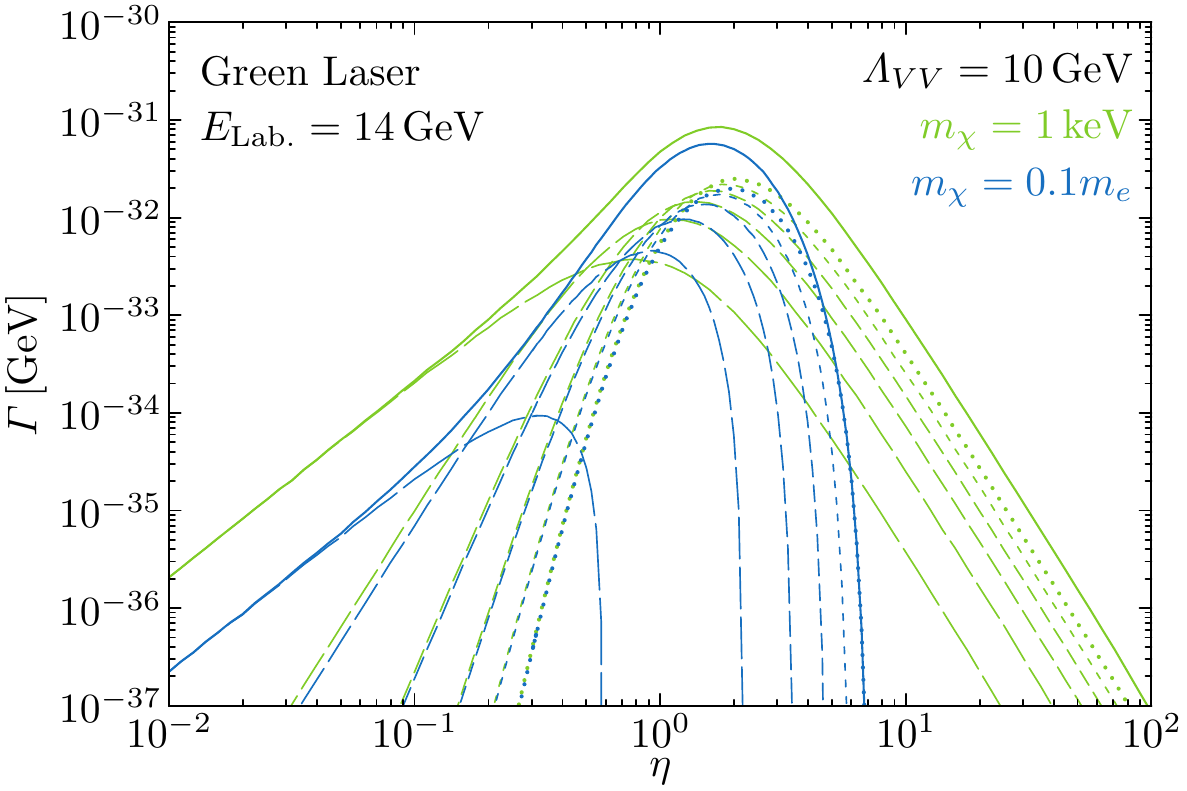}
\includegraphics[width=0.48\textwidth]{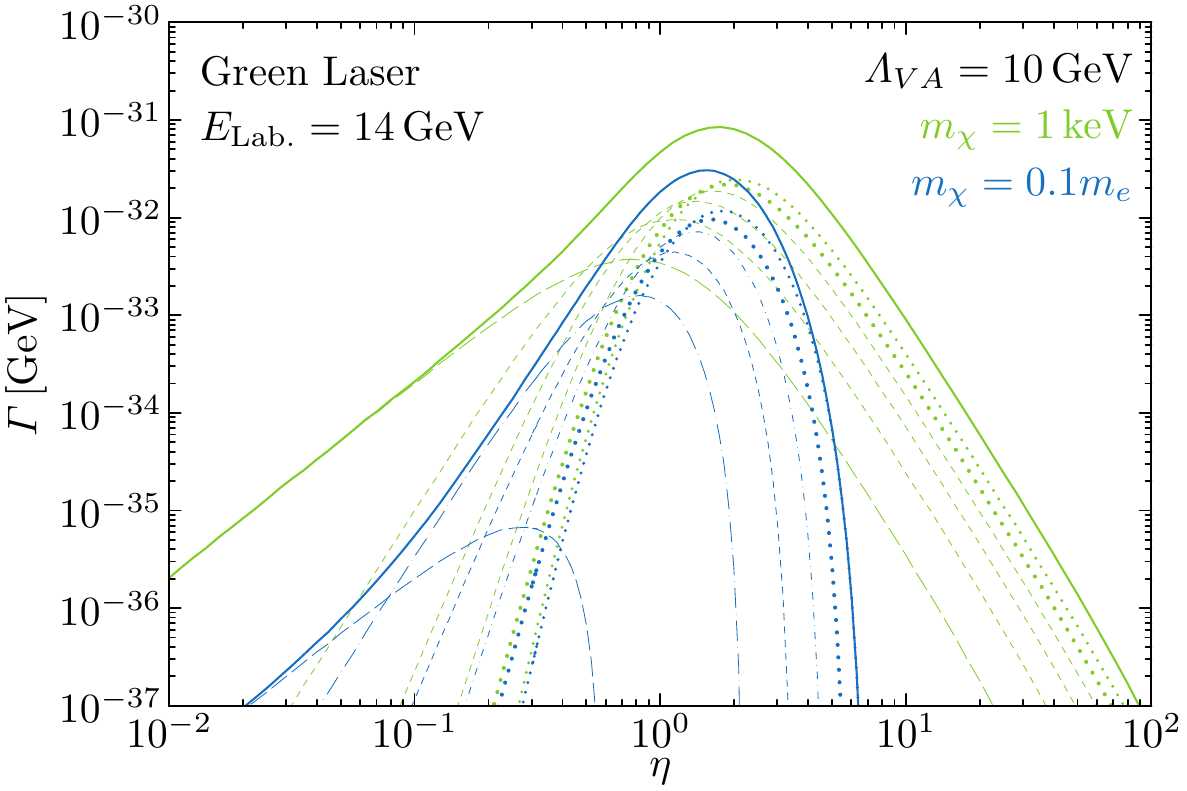}
\\
\includegraphics[width=0.48\textwidth]{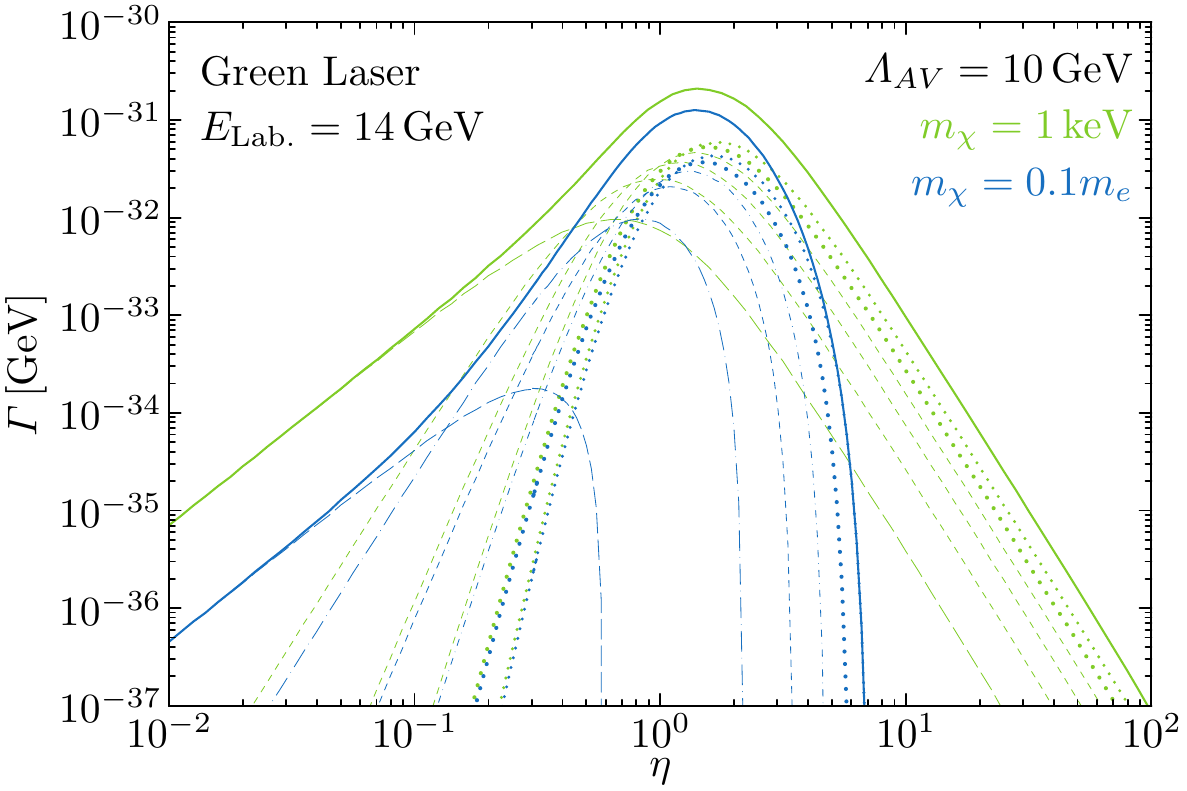}
\includegraphics[width=0.48\textwidth]{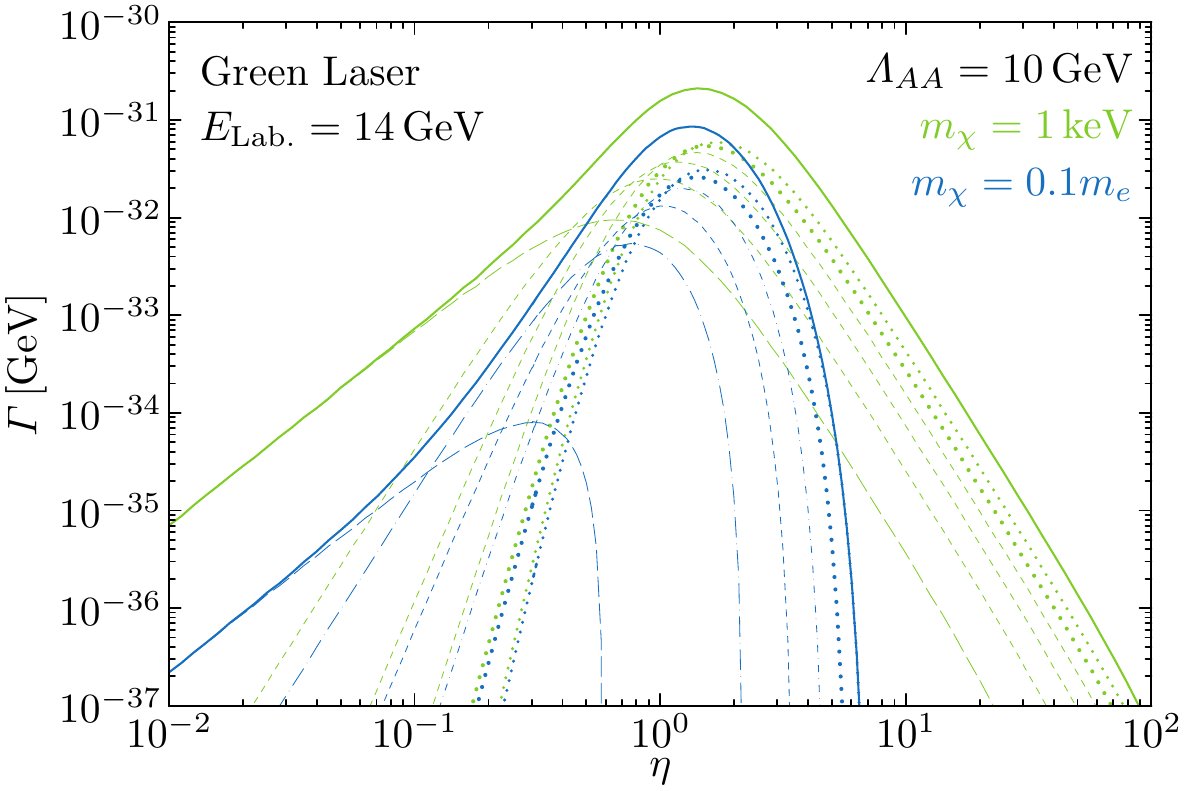}
\caption{The decay width of the laser dressed electron for the operator $\calo_{VV}$ (top-left),
$\calo_{VA}$ (top-right), $\calo_{AV}$ (bottom-left) and $\calo_{AA}$ (bottom-right) as a function of quantity $\eta$, as labeled in Fig.~\ref{fig:SP:Eta}.
}
\label{fig:VA:Eta}
\end{center}
\end{figure}
\begin{figure}[htb!]
\begin{center}
\includegraphics[width=0.48\textwidth]{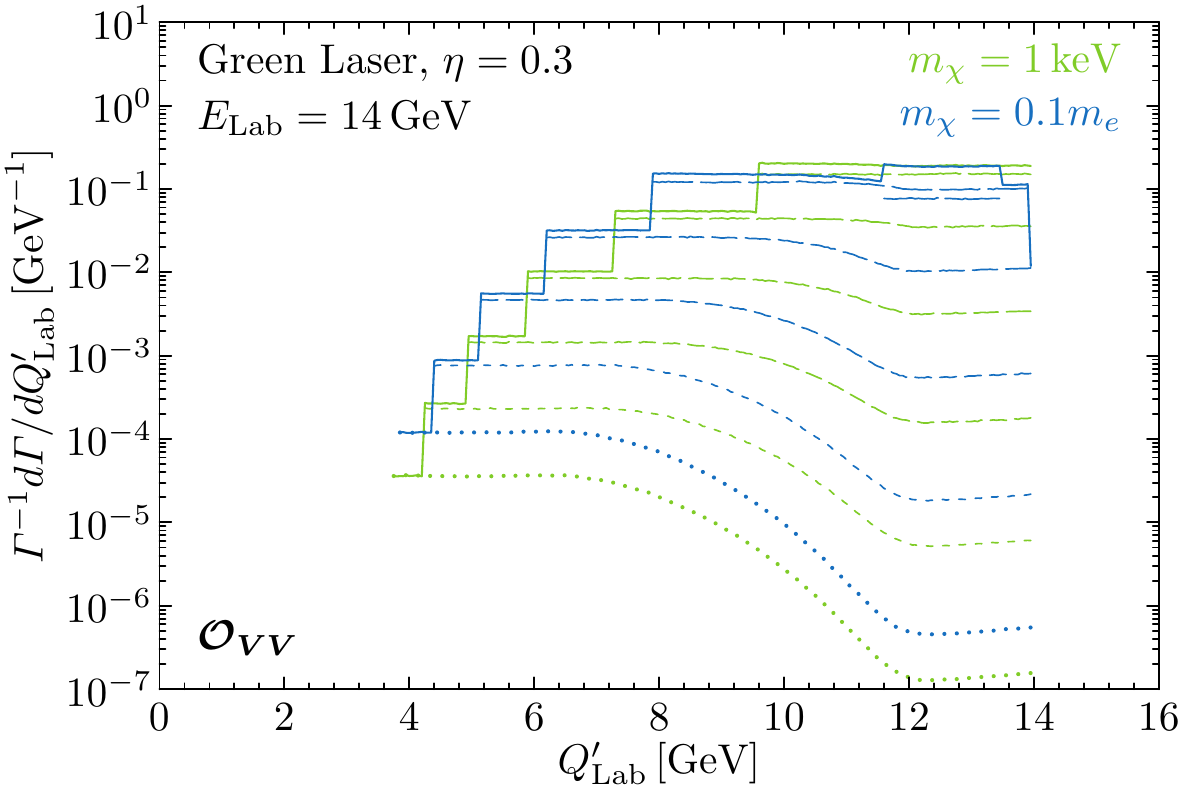}
\includegraphics[width=0.48\textwidth]{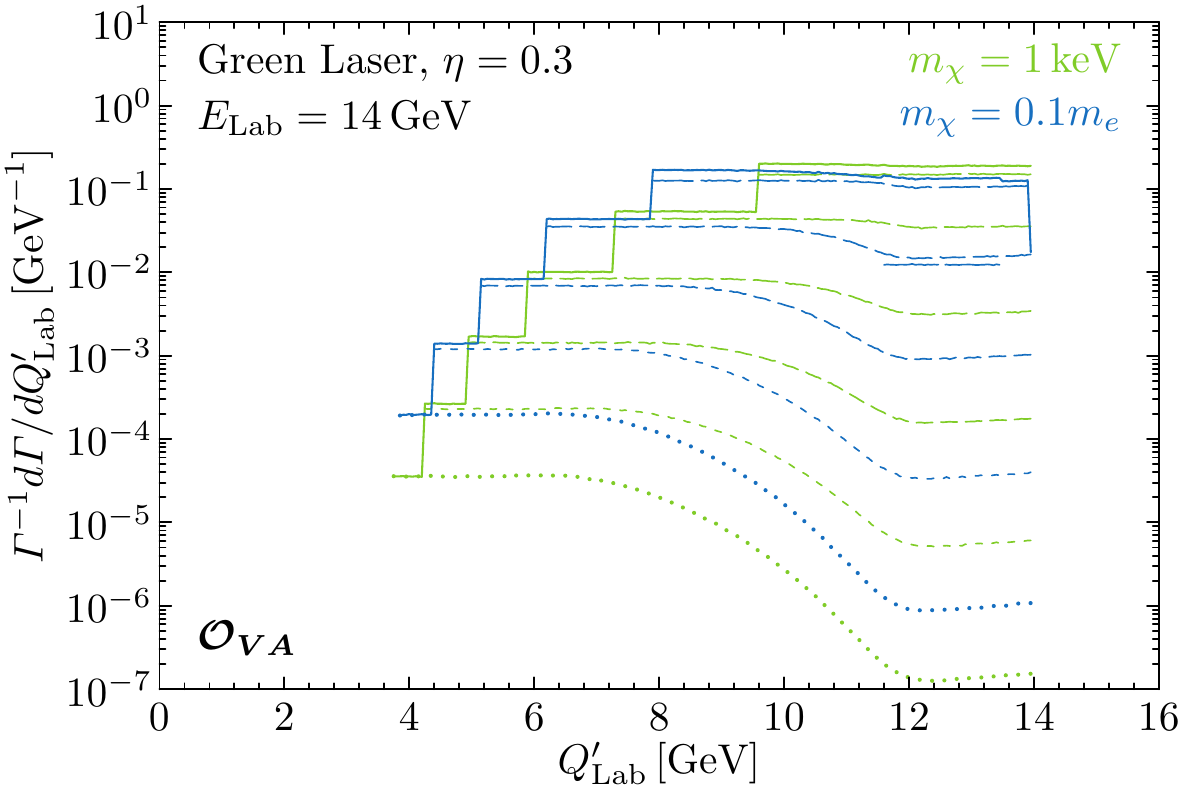}
\\
\includegraphics[width=0.48\textwidth]{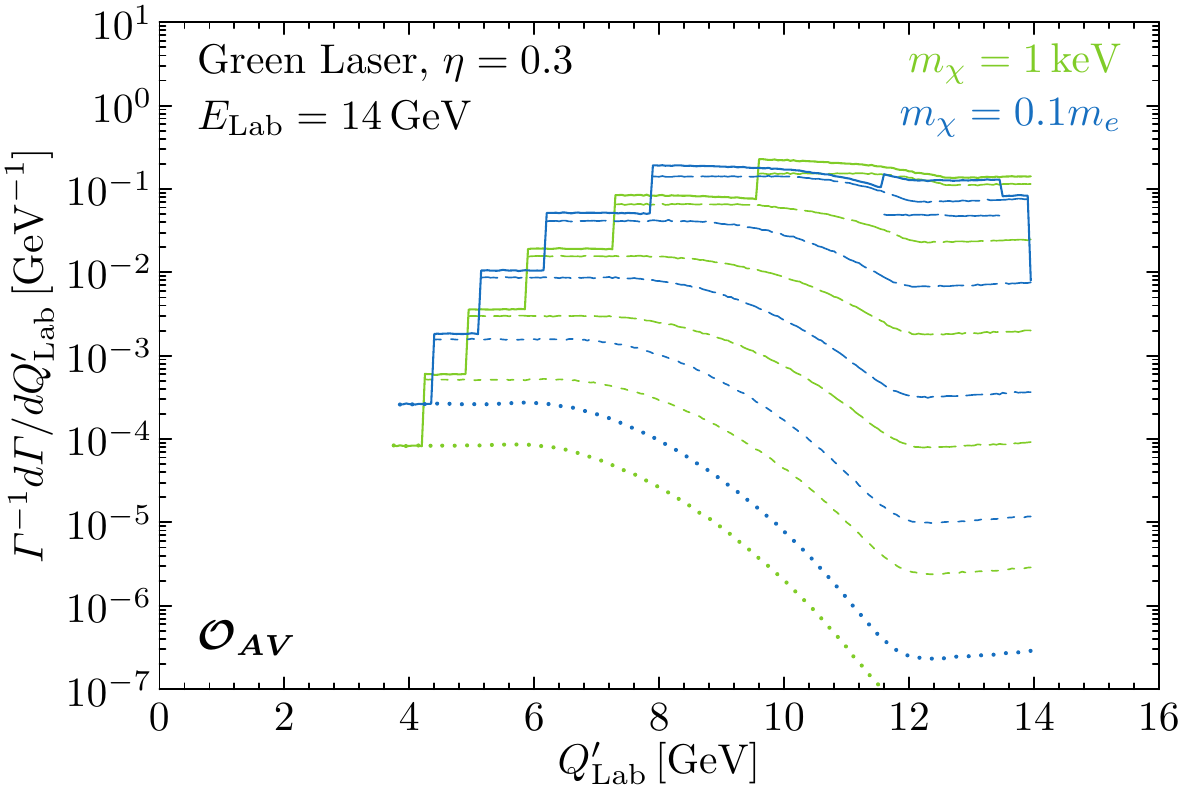}
\includegraphics[width=0.48\textwidth]{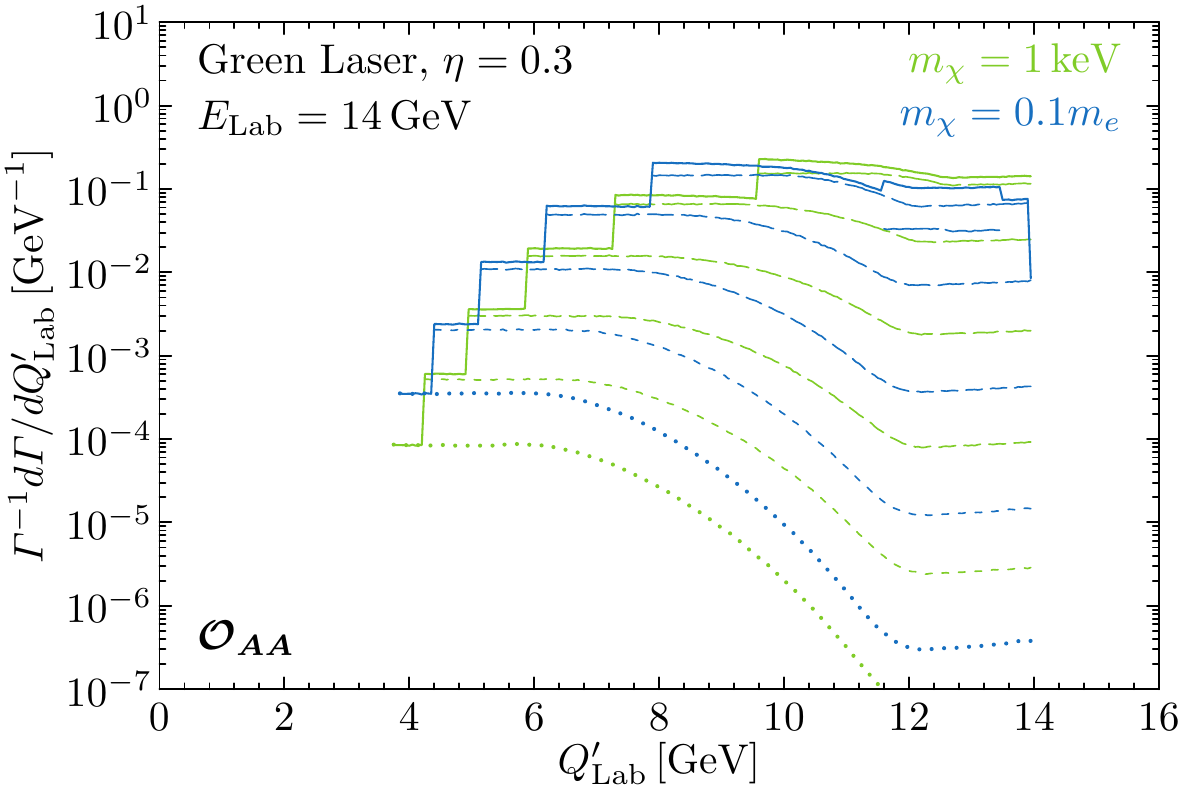}
\caption{The normalized distribution of the decay width with respect to the outgoing electron energy for the operator $\calo_{VV}$ (top-left),
$\calo_{VA}$ (top-right), $\calo_{AV}$ (bottom-left) and $\calo_{AA}$ (bottom-right), as labeled in Fig.~\ref{fig:SP:Eta}.
}
\label{fig:VA:EE}
\end{center}
\end{figure}

Fig.~\ref{fig:VA:Eta} shows the decay widths of the laser dressed electron
for the operator $\calo_{VV}$ (top-left), $\calo_{VA}$ (top-right), $\calo_{AV}$ (bottom-left)
and $\calo_{AA}$ (bottom-right)
as a function of the laser intensity parameter $\eta$.
The results are shown for a fiducial energy scale $\varLambda = 10$ GeV, and a DM mass $m_\chi = 1$ keV (green curves) or $m_\chi = 0.1m_e$ (blue curves).
The total contributions are shown by solid curves,
and the contributions from every branch are denoted by dashed and dotted lines
as shown in the legend of Fig.~\ref{fig:SP:Eta}.
The Fig.~\ref{fig:VA:EE} shows the normalized distribution of the total decay width
with respect to the outgoing electron energy for a dark fermion mass $m_\chi = 1$ keV (green curves) and $m_\chi = 0.1m_e$ (blue curves).
The laser intensity parameter has been set to the typical value from the SLAC experiment \cite{Bamber:1999zt}, namely $\eta=0.3$. The total contributions up to $n=6$ are shown by solid curves,
while the contributions of each branch are also shown by non-solid curves. The features for these operators are similar to those for (pseudo-)scalar operators.

\subsection{virtual photon mediator}

In this subsection, we discuss a mediator model in which the dark fermion $\chi$
interacts with the electrons through dipole couplings,
and the corresponding effective operators are defined in Eq.~(\ref{eq:dipole}).
In this case, the amplitude can be decomposed similar to the (axial-)vector effective operators. However, here we need to insert the metric
decomposition defined in Eq.~(\ref{eq:metric}) twice.
One can easily find that the $n$-th density matrix elements are given as
\bee
\rho_{VX', n}
=
\frac{1}{ m_{k'}^2 } \sum_{\lambda, \lambda' = s, 0, \pm1}
\eta_{\lambda} \eta_{\lambda'}
\calp_{V, n, \lambda\lambda'} \cald_{X', \lambda\lambda'} \,,
\ene
where $1/ m_{k'}^2$ stands for the propagator of the virtual photon.
Furthermore, in the above equation,
we have taken into account that the Lorentz structure on the production side
is always of vector type. Its expressions are already given in Eq.~(\ref{eq:prod:v}).
For the decay part, since all the kinematics of the outgoing DM are integrated out,
the averaged decay density matrix elements are again diagonal.
After tedious calculations, for both the magnetic and electric dipole
operators, we find that the density matrix elements with helicity
$\lambda = s$ always vanish, and the non-vanishing matrix elements
with $\lambda = 0,\pm1$ are given as
\bea
\overline{ \cald_{M, \lambda\lambda' }  }
&=&
\frac{1}{4\pi} m_{k'}^4 \beta_{\chi} \left( 1 - \frac{2}{3}\beta_{\chi}^2 \right) \delta_{\lambda \lambda'}\,,
\quad
\lambda = 0,\,\pm1
\\[3mm]
\overline{ \cald_{E, \lambda\lambda' }  }
&=&
\frac{1}{12} m_{k'}^4 \beta_{\chi}^3 \delta_{\lambda \lambda'}\,,
\quad
\lambda = 0,\,\pm1
\ena
One can see that the magnetic and electric dipole contributions have
different velocity ($\beta_\chi$) dependence.
For ultra-light DM, $\beta_\chi \sim 1$,
the electric dipole contribution is always larger, while the magnetic dipole
contribution becomes larger for heavier DM.
The total contributions are given by integrating over the full range of the
invariant mass $m_{k'}^2$ (and hence the velocity $\beta_\chi$),
but this property is still valid.

\begin{figure}[htb!]
\begin{center}
\includegraphics[width=0.48\textwidth]{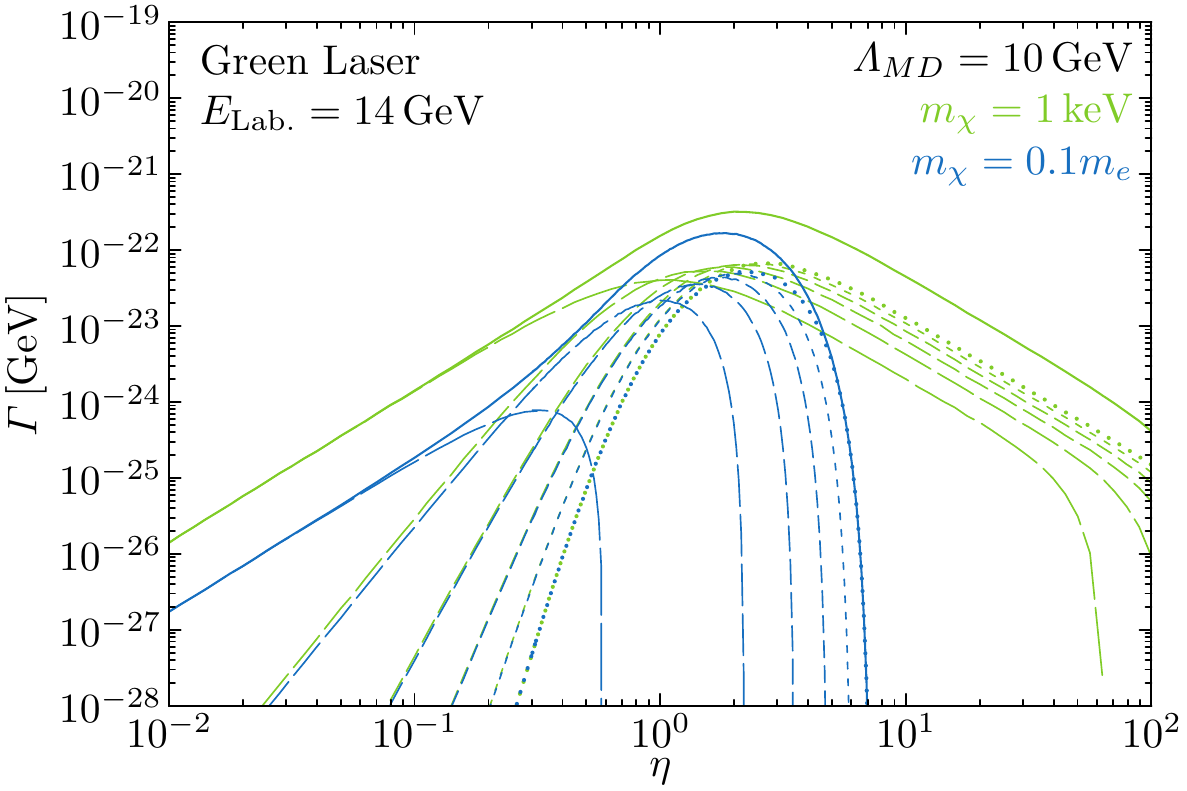}
\includegraphics[width=0.48\textwidth]{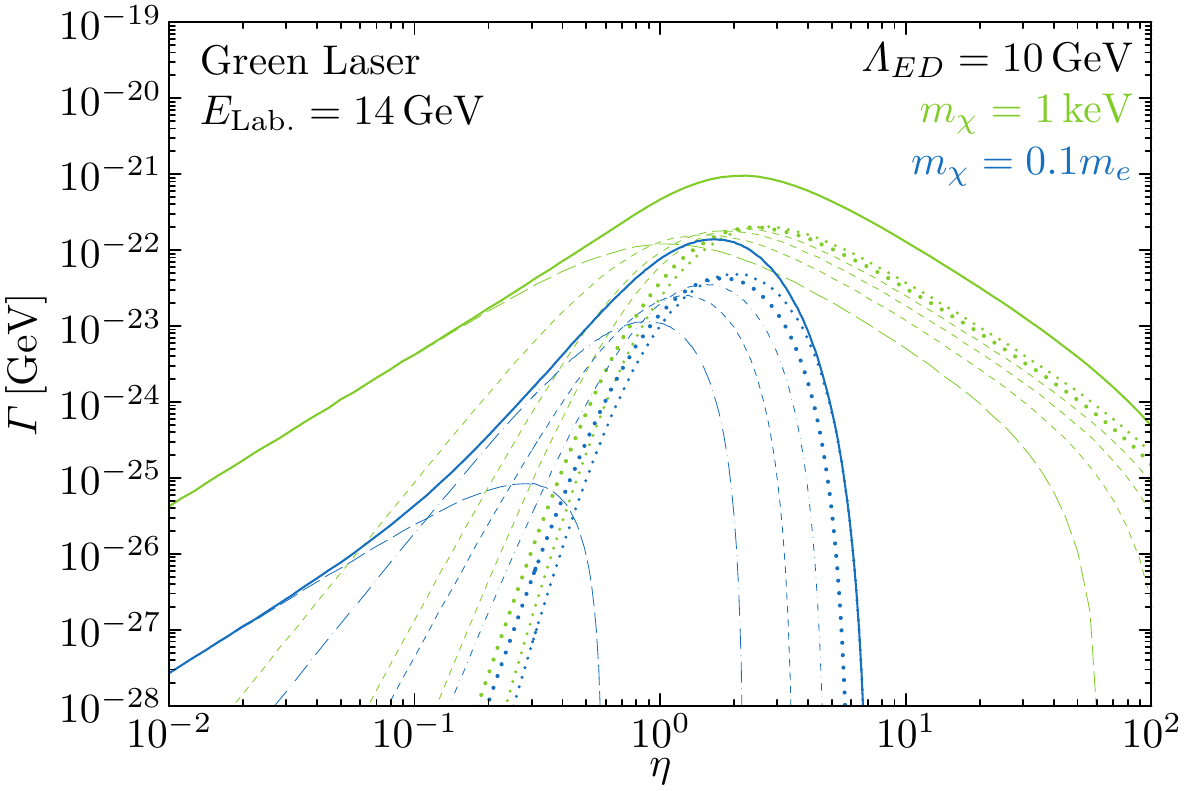}
\caption{
The decay width of the laser dressed electron for the operator $\calo_{MD}$ (left)
and $\calo_{ED}$ (right) as a function of the laser intensity parameter $\eta$.
The meaning of each curve are shown in the legend of the Fig.~\ref{fig:SP:Eta}.
}
\label{fig:EMD:Eta}
\end{center}
\end{figure}
\begin{figure}[htb!]
\begin{center}
\includegraphics[width=0.48\textwidth]{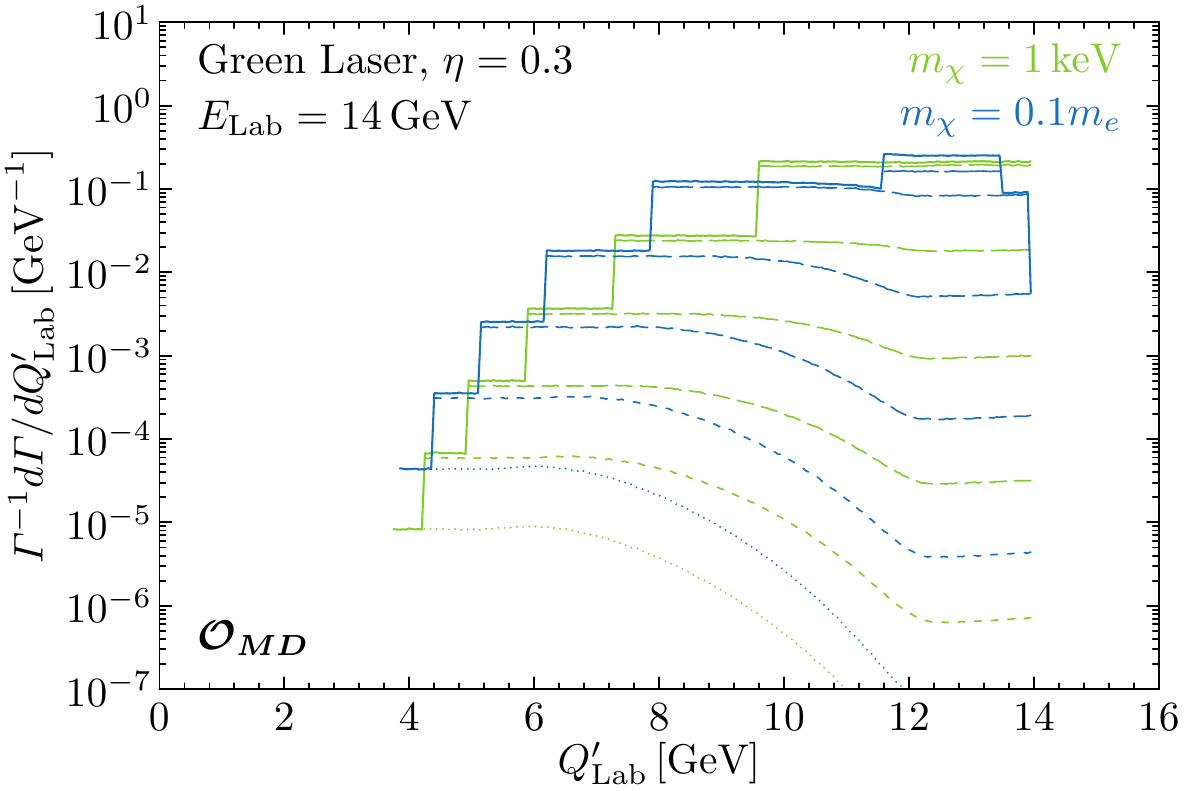}
\includegraphics[width=0.48\textwidth]{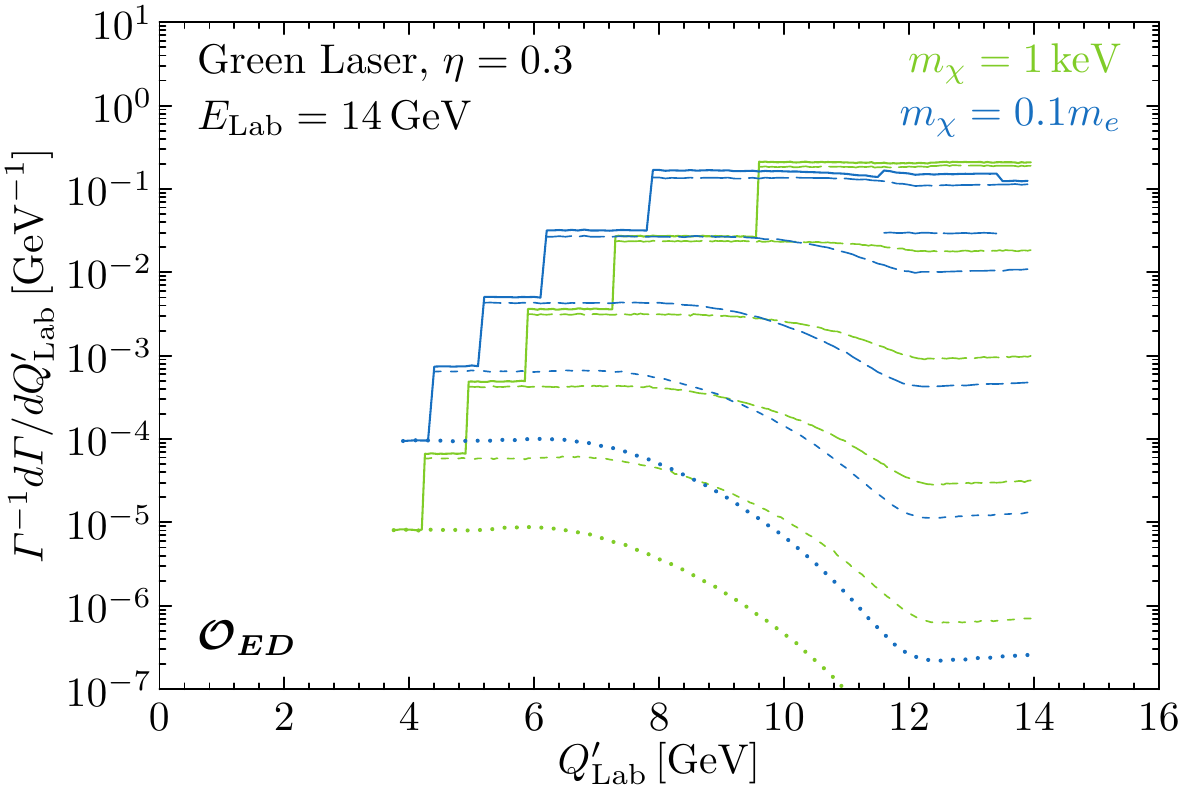}
\caption{
The normalized distribution of the decay width with respect to the outgoing electron energy for the operator $\calo_{MD}$ (left) and $\calo_{ED}$ (right), as labeled in Fig.~\ref{fig:SP:Eta}.
}
\label{fig:EMD:EE}
\end{center}
\end{figure}

Fig.~\ref{fig:EMD:Eta} shows the decay widths of the laser dressed electron
for the operators $\calo_{MD}$ (left) and $\calo_{ED}$ (right)
as a function of the laser intensity parameter $\eta$.
The results are shown for a fiducial energy scale $\varLambda = 10$ GeV, and a DM mass $m_\chi = 1$ keV (green curves) or $m_\chi = 0.1m_e$ (blue curves).
The total contributions are shown by solid curves,
and the contributions from each branch are denoted by dashed and dotted lines
as shown in the legend of Fig.~\ref{fig:SP:Eta}.
The Fig.~\ref{fig:EMD:EE} shows the normalized distribution of the total decay width
with respect to the outgoing electron energy for a dark fermion mass $m_\chi = 1$ keV (green curves) and $m_\chi = 0.1m_e$ (blue curves).
As we have mentioned, while the decay density matrix element
of the electric dipole moment is larger for lighter DM,
the one for the magnetic dipole moment is larger for heavier DM.
This can be seen clearly in Fig.~\ref{fig:EMD:Eta}.
Other features are similar to those for other operators.

\section{Numerical results}
\label{sec:results}

In this section, we show the numerical results of the laser induced Compton scattering to a DM pair. The signal event in our case is composed
of a single electron and missing energy carried away by the DM pair, and becomes
\begin{eqnarray}
N_s = {1\over 2\rho_\omega} \varGamma \cdot \mathcal{L}\;,
\end{eqnarray}
where $\rho_\omega={a^2 \omega_{\rm Lab}\over 4\pi}$ is the laser photon density~\cite{Greiner:1992bv} with ${1\over 2\rho_\omega} \varGamma$ being the scattering cross section~\cite{Greiner:1992bv}, and $\mathcal{L}$ is the integrated luminosity. The luminosity is given by
\begin{eqnarray}
\mathcal{L}=N_e \rho_\omega \ell N_b f t\;,
\end{eqnarray}
where $N_e=1.5\times 10^9$ denotes the number of electrons in a bunch in the electron beam of the European XFEL (EuXFEL) accelerator used for the LUXE experiment~\cite{LUXE:2023crk}, $N_b=2700$ is the number of individual bunches in the beam~\cite{LUXE:2023crk}, $f=1~{\rm Hz}$ is the laser operating frequency~\cite{LUXE:2023crk}, $\ell\simeq 50~{\rm \mu m}$ is the electron pathlength through
the laser focus~\cite{Bamber:1999zt}, and $t=5\times 10^6~{\rm s}$
is the physics data-taking time after taking into account the LUXE data taking efficiency of $75\%$~\cite{LUXE:2023crk}. We obtain $\mathcal{L}\simeq 0.6~{\rm ab}^{-1}$ by choosing $\eta=0.3$ and the above parameters.
This luminosity is too conservative to give sizable constraints on the new physics energy scale.
Higher laser intensity can give larger number density $\rho_\omega$, and hence larger luminosity. However, higher intensity may affect the stability of the laser system. On the other hand,
it is easier to enhance the luminosity by
increasing the number of electrons in each bunch or
the operating frequency. For instance, the BESIII experiment can achieve an operating frequency above 1\,kHz~\cite{BESIII:2009fln}. Thus, a much higher luminosity can be expected.

\begin{figure}[htb!]
\begin{center}
\includegraphics[width=0.48\textwidth]{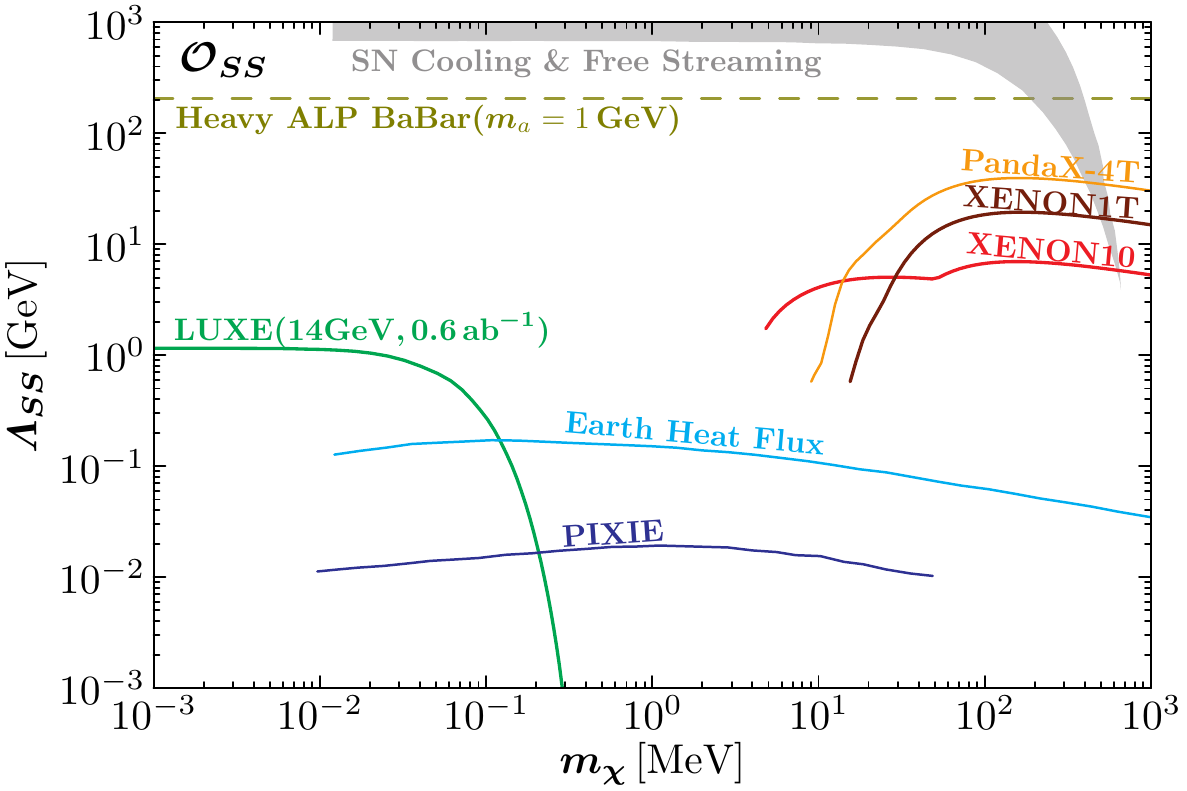}
\includegraphics[width=0.48\textwidth]{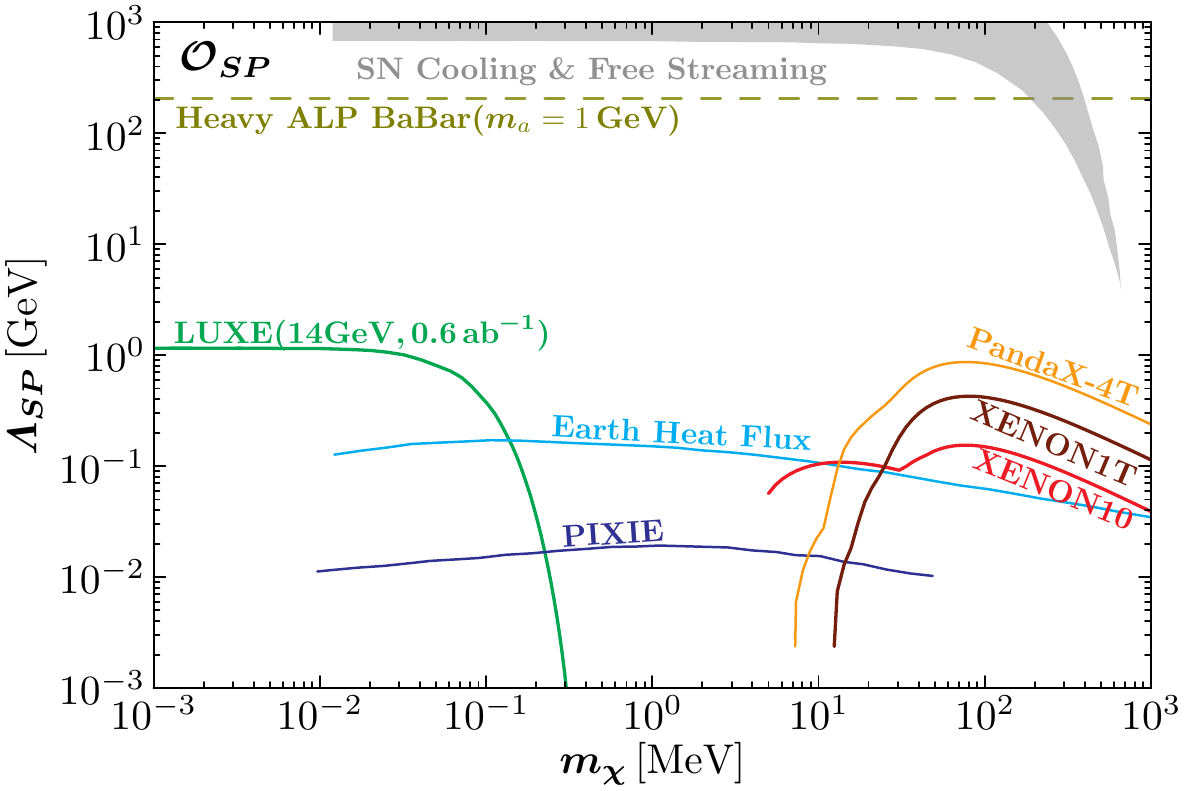}
\\
\includegraphics[width=0.48\textwidth]{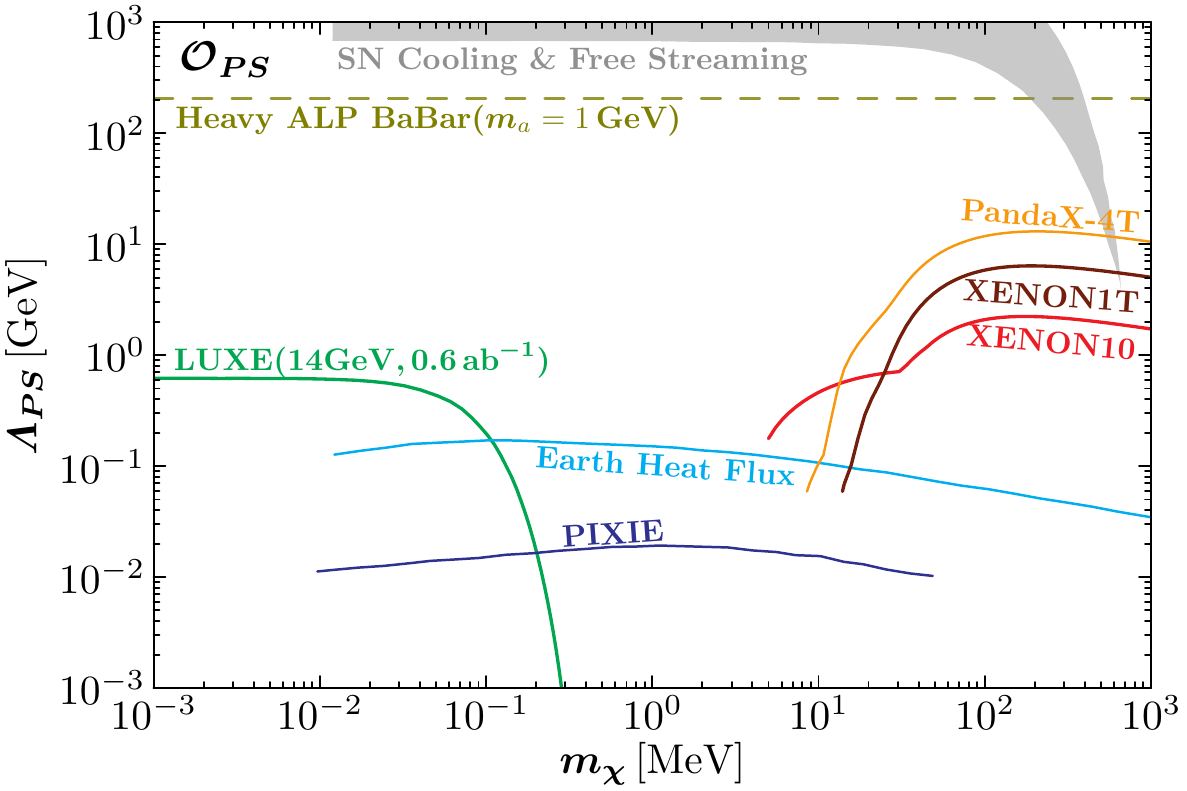}
\includegraphics[width=0.48\textwidth]{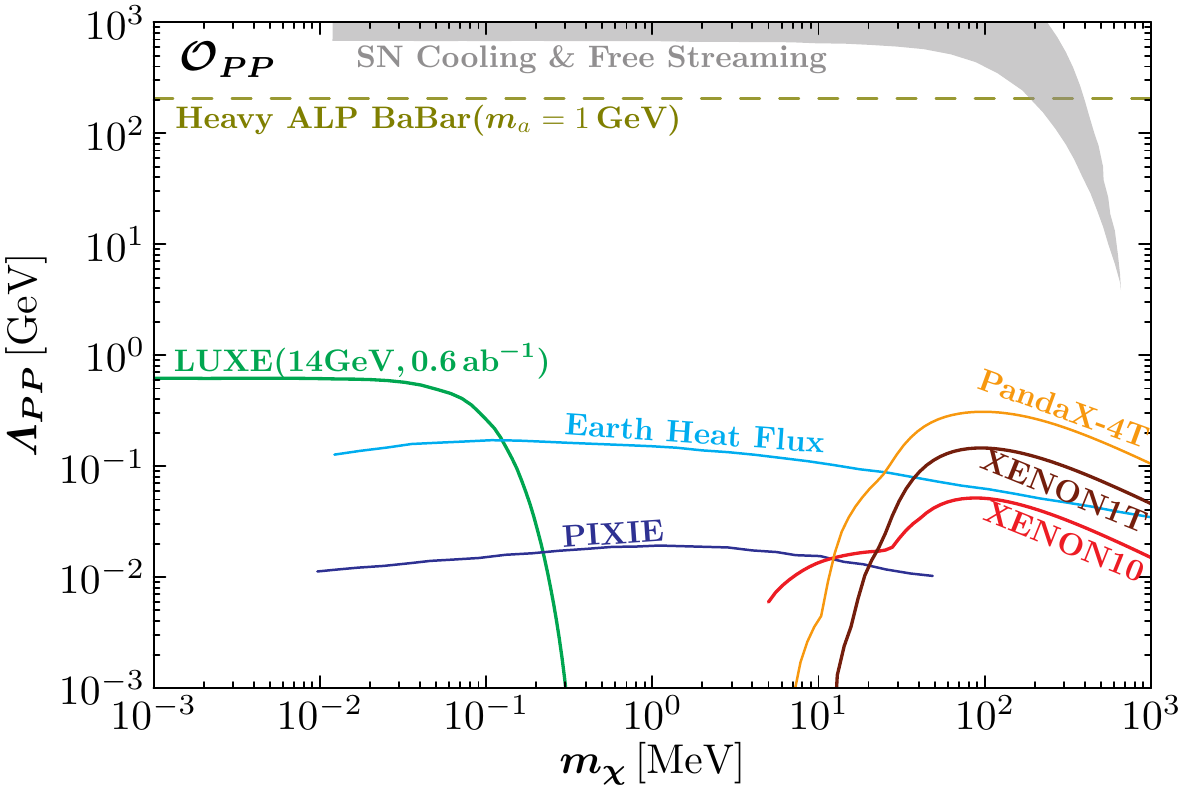}
\caption{Expected limits (green curve) on $\Lambda_{SS}$ (top-left), $\Lambda_{SP}$ (top-right), $\Lambda_{PS}$ (bottom-left) and $\Lambda_{PP}$ (bottom-right). The limits~\cite{Liang:2024ecw} from PandaX-4T~\cite{PandaX:2022xqx}, XENON1T~\cite{XENON:2019gfn}, XENON10~\cite{Essig:2017kqs}, and astrophysical bounds~\cite{Guha:2018mli} such as SN cooling and free-streaming~\cite{Arnett:1989tnf}, heat flux of Earth~\cite{Chauhan:2016joa} and PIXIE~\cite{Ali-Haimoud:2015pwa} are also shown for comparison.
The translated exclusion limit is given by the constraint on heavy ALP with $m_a=1$ GeV at BaBar~\cite{Bauer:2017ris,Bauer:2018uxu,Liu:2023bby}.
}
\label{fig:CS:SP}
\end{center}
\end{figure}

\begin{figure}[htb!]
\begin{center}
\includegraphics[width=0.48\textwidth]{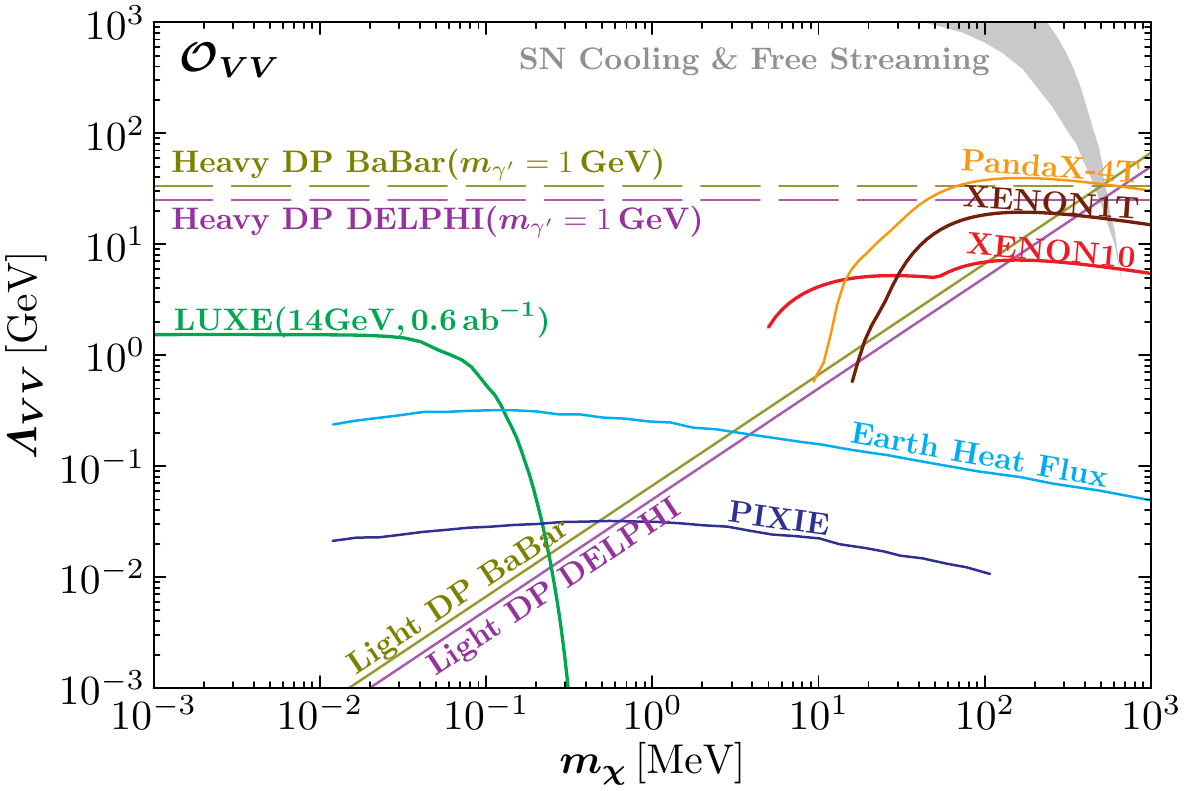}
\includegraphics[width=0.48\textwidth]{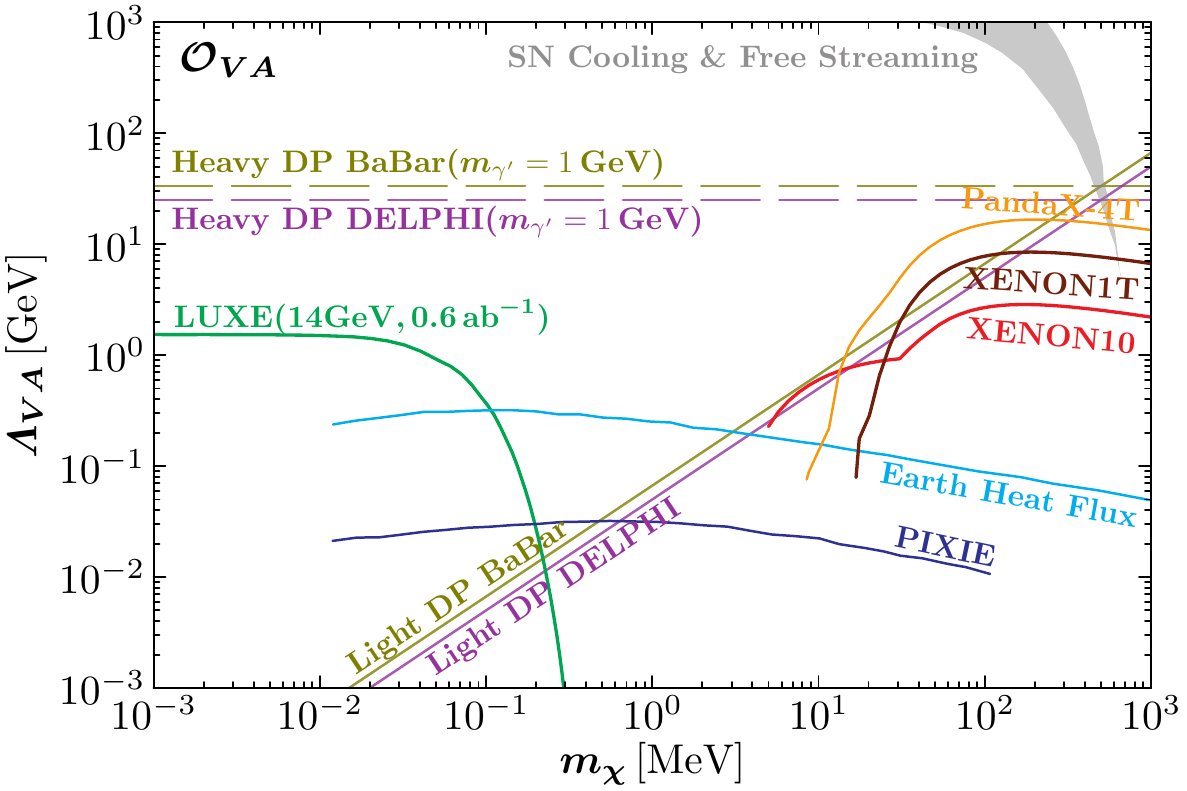}
\\
\includegraphics[width=0.48\textwidth]{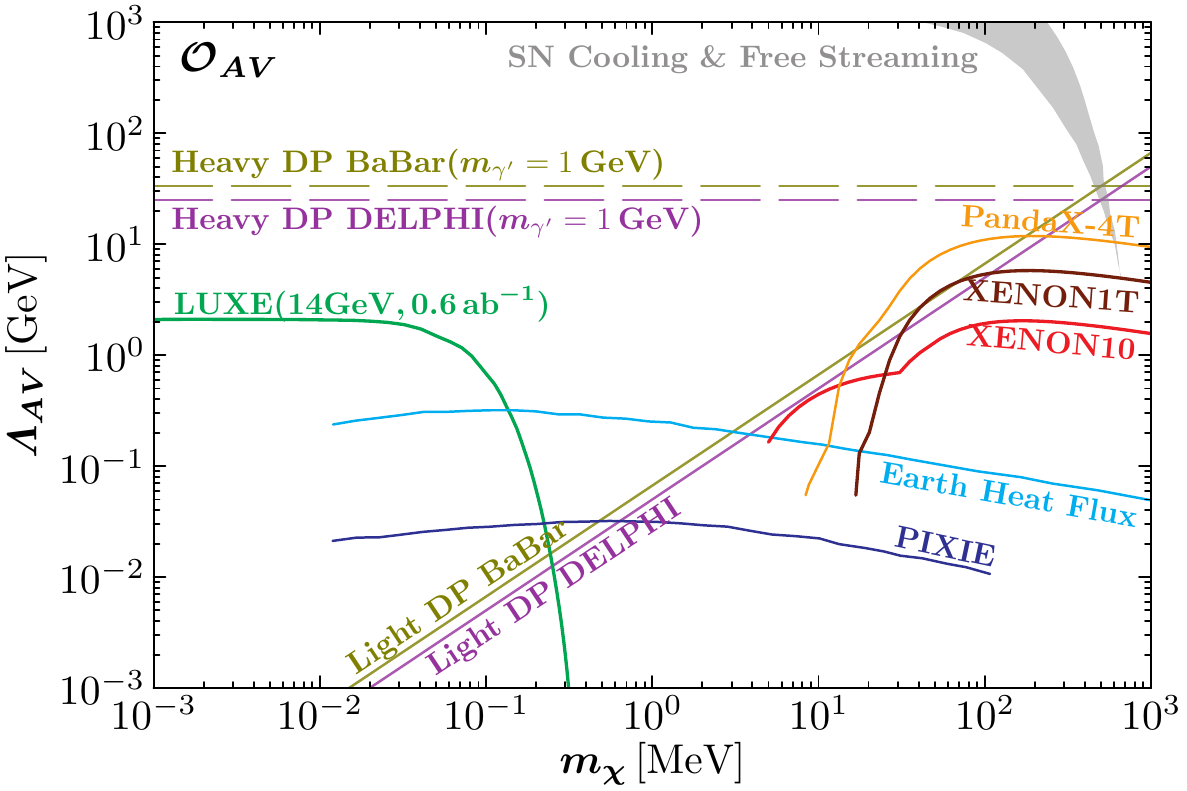}
\includegraphics[width=0.48\textwidth]{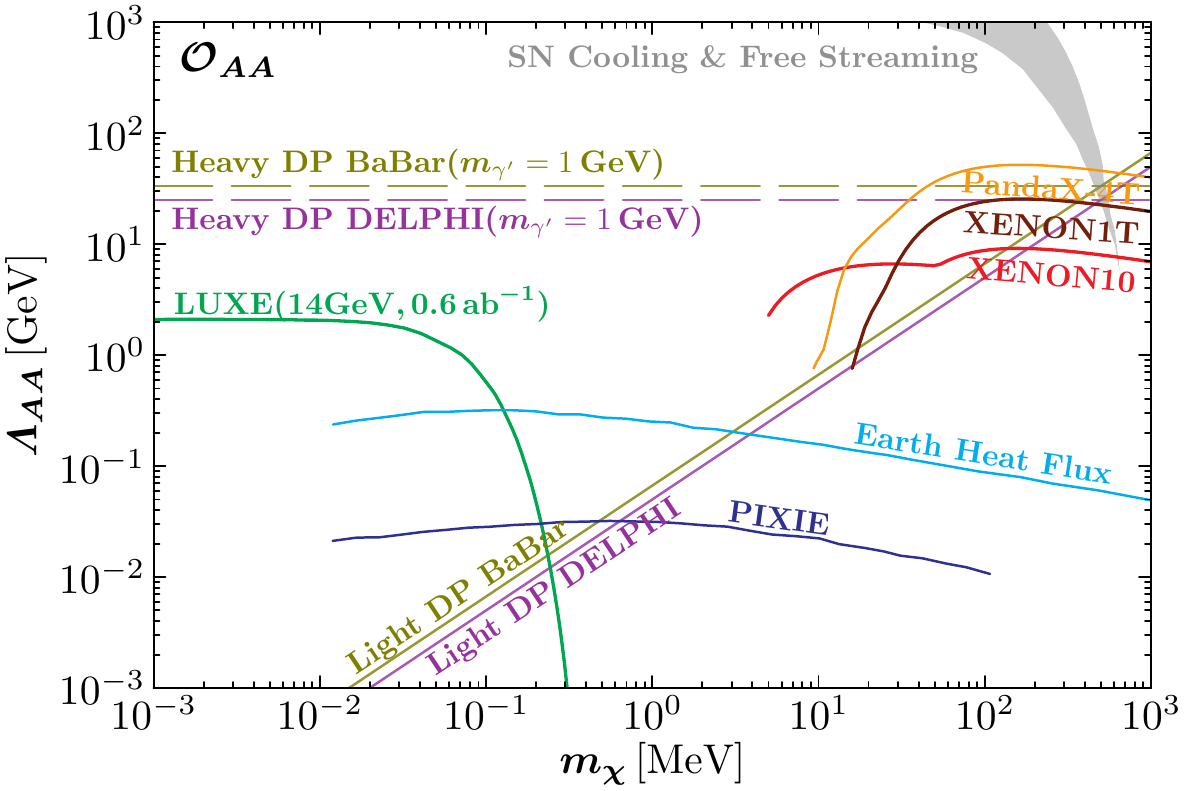}
\caption{Expected limits (green curve) on $\Lambda_{VV}$ (top-left), $\Lambda_{VA}$ (top-right), $\Lambda_{AV}$ (bottom-left) and $\Lambda_{AA}$ (bottom-right). The limits~\cite{Liang:2024ecw} from PandaX-4T~\cite{PandaX:2022xqx}, XENON1T~\cite{XENON:2019gfn}, XENON10~\cite{Essig:2017kqs}, and astrophysical bounds~\cite{Guha:2018mli} such as SN cooling and free-streaming~\cite{Arnett:1989tnf}, heat flux of Earth~\cite{Chauhan:2016joa} and PIXIE~\cite{Ali-Haimoud:2015pwa} are also shown for comparison.
The translated exclusion limits are given by the constraints on light DP or heavy DP with $m_{\gamma'}=1$ GeV at both BaBar~\cite{BaBar:2017tiz} and DELPHI~\cite{Ma:2022cto}.
}
\label{fig:CS:VA}
\end{center}
\end{figure}

\begin{figure}[htb!]
\begin{center}
\includegraphics[width=0.48\textwidth]{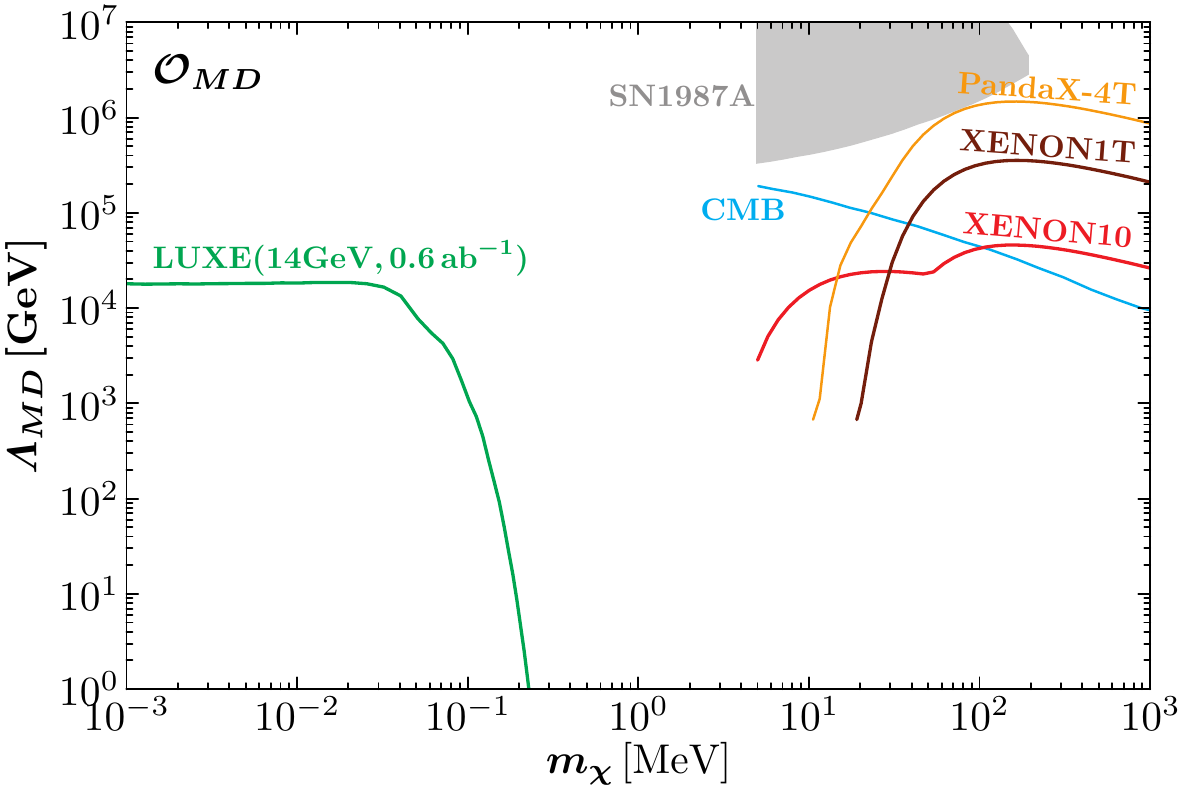}
\includegraphics[width=0.48\textwidth]{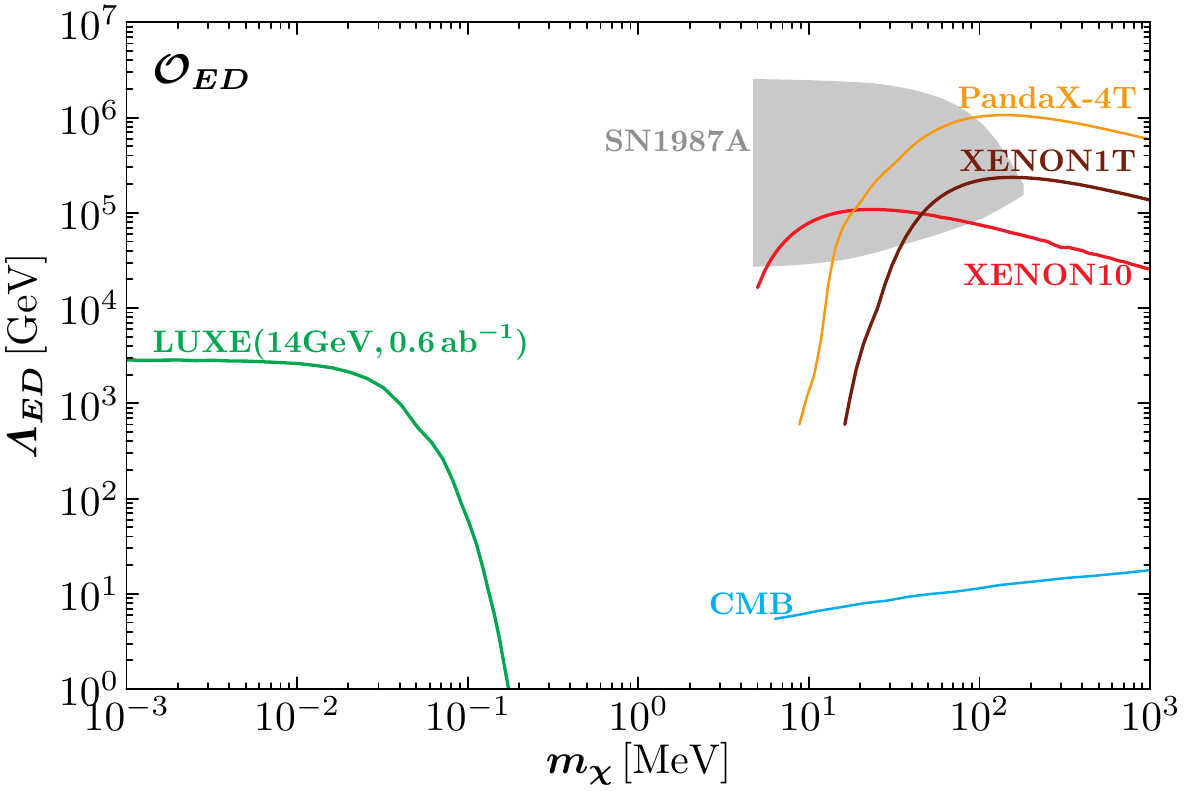}
\caption{Expected limits (green curve) on $\Lambda_{MD}$ (left) and $\Lambda_{ED}$ (right). The limits~\cite{Liang:2024ecw} from PandaX-4T~\cite{PandaX:2022xqx}, XENON1T~\cite{XENON:2019gfn}, XENON10~\cite{Essig:2017kqs}, and astrophysical bounds~\cite{Liang:2024tef} such as SN and CMB are also shown for comparison.
}
\label{fig:CS:ME}
\end{center}
\end{figure}

We assume the observed signal event number to be $N_s=10$ with $E_{\rm Lab}=14~{\rm GeV}$, $\mathcal{L}=0.6~{\rm ab}^{-1}$ and obtain the projected limits on the energy scale of DM EFT as shown in Figs.~\ref{fig:CS:SP}, \ref{fig:CS:VA} and \ref{fig:CS:ME}. The reachable bound for the UV energy scale is about 1 GeV for the dimension-6 operators and $m_\chi<1~{\rm MeV}$. For the dimension-5 magnetic (electric) dipole operator, the expected bound can reach as high as $2\times 10^4~{\rm GeV}$ ($3\times 10^3~{\rm GeV}$). For comparison, the limits~\cite{Liang:2024ecw} from direct detection experiments PandaX-4T~\cite{PandaX:2022xqx}, XENON1T~\cite{XENON:2019gfn} and XENON10~\cite{Essig:2017kqs} are also shown. Although they can reach a higher UV energy scale, these direct detection experiments and others~\cite{CDEX:2022kcd,DAMIC-M:2023gxo,SENSEI:2023zdf,SENSEI:2024yyt} are difficult to probe the regime of DM mass smaller than 1 MeV.

We also compare our results with other constraints on the low-mass regime $m_\chi < 1$ MeV. We cite the astrophysical bounds on dimension-6 operators from Ref.~\cite{Guha:2018mli}, as shown in Figs.~\ref{fig:CS:SP} and \ref{fig:CS:VA}. For scalar/pseudo-scalar (vector/axial-vector) operators~\footnote{Ref.~\cite{Guha:2018mli} performed the analysis for the sum of scalar and pseudo-scalar (or vector and axial-vector) operators with universal coefficients.}, the region between 0.63 TeV and 2.78 TeV (1.12 TeV and 2.99 TeV) has been excluded by supernova (SN) cooling and free-streaming bounds~\cite{Arnett:1989tnf} but is beyond the reachable bound of laser induced process. More recent SN bounds are
obtained based on the full radial thermodynamical profiles from state-of-the-art simulations~\cite{Manzari:2023gkt}, in particular in the trapping regime. The scalar/pseudo-scalar (vector/axial-vector) operators are excluded in the range of 0.07 - 5.4 TeV (0.11 - 5.9 TeV) in the limit of $m_\chi=0$. The region below the cyan line is excluded by the constraint from the internal heat flux of Earth~\cite{Chauhan:2016joa}. The limit from PIXIE~\cite{Ali-Haimoud:2015pwa} is shown by the blue line. These limits are less stringent than those from laser induced Compton scattering. We also translate the upper limits on the coupling of mediator dark photon (DP with kinetic mixing $\epsilon$ and mass $m_{\gamma'}$) or axion-like particle (ALP with coupling to electron $g_{ae}$ and mass $m_a$) from other collider experiments to relevant EFT energy scale. These limits depend on the comparison between the mass of DP or ALP and the transfer momentum. For illustration, the limit on EFT energy scale of vector/axial-vector operators becomes
\begin{eqnarray}
\left\{
  \begin{array}{ll}
    \Lambda>m_{\gamma'}/\sqrt{\epsilon_{\rm limit}}, & \hbox{when $m_{\gamma'}^2\gg (p_\chi+p_{\overline{\chi}})^2$;} \\
    \Lambda>2m_\chi/\sqrt{\epsilon_{\rm limit}}, & \hbox{when $m_{\gamma'}^2\ll (p_\chi+p_{\overline{\chi}})^2$.}
  \end{array}
\right.
\end{eqnarray}
For scalar/pseudo-scalar operators, one can just replace $\epsilon$ and $m_{\gamma'}$ by ALP-electron coupling $g_{ae}$ and ALP mass $m_a$, respectively. In Fig.~\ref{fig:CS:VA}, we show the representative bound on EFT energy scale of vector/axial-vector operators from the second case. One can see that a part of parameter space is excluded but it is weaker than the projected limit from our laser induced process. However, if the mediator mass is large enough as shown in the first case, the collider constraint is rather strong and excludes laser sensitivity parameter space. For example, taking $m_{\gamma'}=1$ GeV for DP ($m_a=1$ GeV for ALP), the exclusion limit from BaBar experiment~\cite{BaBar:2017tiz,Bauer:2017ris,Bauer:2018uxu,Liu:2023bby} becomes $\Lambda\sim 32$ (200) GeV as shown in Fig.~\ref{fig:CS:VA} (Fig.~\ref{fig:CS:SP}).
In addition, the projected bound from mono-photon channel at high-energy $e^+e^-$ collider would be as strong as $\sim 4$ TeV~\cite{Kundu:2021cmo}. However, compared with the c.m. collision energy, the obtained lower limit of UV energy scale is not large enough to ensure the EFT validity. We thus do not show it here and refer readers to the relevant references~\cite{Barman:2021hhg,Kundu:2021cmo}.
For the two dimension-5 dipole operators, Ref.~\cite{Liang:2024tef} derived the relation between them and the dimension-6 quark tensor operators, and obtained the constraints from long-distance contribution. Under the flavor symmetric case, we translate their limits on the quark tensor operators to those for dimension-5 operators. Fig.~\ref{fig:CS:ME} includes the constraints from SN and cosmic microwave background (CMB) observations.

\section{Conclusion}
\label{sec:Con}

The collision of
an ultra-relativistic electron beam with a laser pulse in an intense electromagnetic field induces the decay of the electron. The precision measurement of this nonlinear Compton scattering catalyzes the studies of nonlinear QED and the search for new physics in terrestrial experiments. The high-intensity facility with an intense electromagnetic field may compensate for the shortcomings
of other experiments.

In this work, we propose to search for the nonlinear Compton scattering producing a pair of DM particles in the presence of a high-intensity laser field. We take into account the Dirac-type fermionic DM in a leptophilic scenario and the DM-electron interactions in the framework of effective field theory. The rates of electron decay to a fermionic DM pair are calculated for effective DM operators of different bilinear products. We find that
\begin{itemize}
\item The absorption of multiple laser photons significantly increases the width of electron decay to a DM pair, making it several orders of magnitude larger than the result for $n=1$.
\item As the number $n$ increases, the energy of the outgoing electron exhibits continuous edges in lower energy region, and the differential decay width correspondingly decreases.
\item The reachable bound on the UV energy scale is about 1 GeV for the dimension-6 operators in DM EFT, under a conservative luminosity of $\mathcal{L}=0.6~{\rm ab}^{-1}$. For the dimension-5 dipole operators, the expected bound can reach as high as $10^3-10^4~{\rm GeV}$.
\item The laser induced process provides a complementary search for DM particles with $m_\chi<1$ MeV, compared with direct detection experiments.
\end{itemize}

\acknowledgments

T.~L. is supported by the National Natural Science Foundation of China (Grant No. 12375096, 12035008, 11975129). K. M. was supported by the Natural Science Basic Research Program of Shaanxi (Program No. 2023-JC-YB-041), and the Innovation Capability Support Program of Shaanxi (Program No. 2021KJXX-47).

\bibliographystyle{JHEP}
\bibliography{refs}

\end{document}